\documentclass{aa}

\usepackage{graphicx}

\usepackage{txfonts}

\usepackage{float}
\usepackage[utf8]{inputenc}
\usepackage[T1]{fontenc}
\usepackage{textcomp}
\usepackage{newunicodechar}
\usepackage{multirow}
\usepackage{threeparttablex}

\DeclareUnicodeCharacter{2212}{-}   
\DeclareUnicodeCharacter{2013}{--} 
\DeclareUnicodeCharacter{2014}{---} 
\DeclareUnicodeCharacter{00A0}{~} 
\DeclareUnicodeCharacter{202F}{~}

\usepackage{tikz}
\usetikzlibrary{shapes.geometric, arrows}

\tikzstyle{startstop} = [rectangle, rounded corners, minimum width=3cm, minimum height=1cm,text centered, draw=black, fill=red!30]
\tikzstyle{process} = [rectangle, minimum width=3cm, minimum height=1cm, text centered, draw=black, fill=blue!20]
\tikzstyle{decision} = [diamond, minimum width=3cm, minimum height=1cm, text centered, draw=black, fill=green!30]
\tikzstyle{arrow} = [thick,->,>=stealth]

\usepackage[hidelinks]{hyperref}

\usepackage{threeparttable}
\usepackage{multicol}
\usepackage{pdflscape}
\usepackage{longtable}

\usepackage[switch]{lineno}

\defcitealias{2025MNRAS.537.1015T}{CGTP25}

\begin{document} 

\titlerunning{Ca-strong SNe}
\title{Peak `Nebular' Emission and Early Flux Excesses in Ca-strong Supernovae}

\authorrunning{C.-G. Touchard-Paxton et al.}
\author{C.-G.~Touchard-Paxton\inst{1},
        K.~Maguire\inst{1},
        C.~Frohmaier\inst{2},
        W.V.~Jacobson-Gal\'{a}n\thanks{NASA Hubble Fellow}\inst{3},
        M.~Pursiainen\inst{4},
        C.~Angus\inst{5},
        J.~Sollerman\inst{6},
        A.~Polin\inst{7},
        J.H.~Terwel\inst{1},
        R.~Seth\inst{1},
        C.~O'Donnell\inst{1},
        S.J~Smartt\inst{8,5},
        S.~Srivastav\inst{8}, 
        A.~Bochenek\inst{9},
        T.~de~Boer\inst{10},
        K.K.~Das\inst{11},
        C.~Fremling\inst{12,13},
        A.C.~Gordon\inst{14,15},
        M.J.~Graham\inst{13},
        M.E.~Huber\inst{10},
        C.-C.~Lin\inst{10},
        C.~Liu\inst{14,15,16},
        F.J.~Masci\inst{17},
        P.~Minguez\inst{10},
        G.S.H.~Paek\inst{10},
        J.~Purdum\inst{12},
        B.~Rusholme\inst{16},
        K.~Smith\inst{8},
        R.~Smith\inst{12},
        R.~Wainscoat\inst{10}
        }

\institute{School of Physics, Trinity College Dublin, the University of Dublin, College Green, Dublin 2, Ireland. \email{toucharc@tcd.ie} 
           \and
           Institute of Cosmology and Gravitation, University of Portsmouth, Portsmouth, PO1 3FX, UK
           \and 
           Cahill Center for Astrophysics, California Institute of Technology, MC 249-17, 1216 E California Boulevard, Pasadena, CA, 91125, USA
           \and
           Department of Physics, University of Warwick, Gibbet Hill Road, Coventry CV4 7AL, UK
           \and
           Astrophysics Research Centre, School of Mathematics and Physics, Queen’s University Belfast, Belfast BT7 1NN, UK
           \and
           The Oskar Klein Centre, Department of Astronomy, Stockholm University, AlbaNova, SE-10691 Stockholm, Sweden
           \and
           Department of Physics and Astronomy, Purdue University, 525 Northwestern Avenue, West Lafayette, IN 47907, USA
           \and
           Department of Physics, University of Oxford, Keble Road, Oxford, OX1 3RH, UK
           \and
           Astrophysics Research Institute, Liverpool John Moores University, 146 Brownlow Hill, Liverpool L3 5RF, UK
           \and
           Institute for Astronomy, University of Hawai'i, 2680 Woodlawn Drive, Honolulu, HI 96822, USA
           \and
           Cahill Center for Astrophysics, California Institute of Technology, MC 249-17, 1200 E California Boulevard, Pasadena, CA, 91125, USA
           \and
           Caltech Optical Observatories, California Institute of Technology, Pasadena, CA 91125, USA
           \and
           Division of Physics, Mathematics and Astronomy, California Institute of Technology, Pasadena, CA 91125, USA
           \and
           Center for Interdisciplinary Exploration and Research in Astrophysics (CIERA) 
           \and 
           Department of Physics and Astronomy, Northwestern University, Evanston, IL 60208, USA
           \and
           NSF-Simons AI Institute for the Sky (SkAI), 172 E. Chestnut St., Chicago, IL 60611, USA
           \and
           IPAC, California Institute of Technology, 1200 E. California Blvd, Pasadena, CA 91125, USA
           }

\abstract
    {Calcium-strong supernovae (Ca-strong SNe) are a class of thermonuclear transient characterised by rapid photometric evolution (rise times of $\leq16$~d) and nebular-phase spectra dominated by strong [Ca~II]~\(\lambda\lambda\)7291,7324 emission, in comparison to the strength of [O~I]~\(\lambda\lambda\)6300,6364 emission. Despite their unique spectroscopic signatures and association with remote environments at large physical offsets from their host galaxy, their origins remain uncertain. Using new discoveries from surveys such as the Zwicky Transient Facility, supplemented by archival data, we compile a sample of 46 Ca-strong SNe, including 8 newly-classified objects -- the largest sample presented to date. We present a comprehensive optical photometric and spectroscopic study of Ca-strong SNe, aimed at better constraining their diversity, progenitor systems, and explosion mechanisms.  This includes Gaussian decomposition of [Ca~II] and [O~I] nebular emission lines to measure bulk velocities and line profile evolution from peak brightness through to deep nebular-phase epochs. Of the 35 Ca-strong SNe with spectra near peak light, we find that 32 objects exhibit significant [Ca~II] emission within 10 days of peak. Forbidden [Ca~II] emission is typically expected to emerge after the ejecta have expanded, cooled, and reached a sufficiently low density for forbidden transitions to occur. The prevalence of these early features in Ca-strong SNe, however, remains unexplained by current explosion models of these objects. In several objects, we identify complex, multi-component line profiles that suggest a transition between two distinct velocity regimes. These transitional features, combined with the common presence of early forbidden emission, lead us to favour a scenario in which extended material surrounds the progenitor and pre-dates the SN itself. Our results suggest that the diverse spectroscopic behaviours of Ca-strong SNe are consistent with white dwarfs exploding in extended environments polluted by prior activity of the progenitor system.}

   \keywords{supernovae: general -- stars: binaries -- novae -- supernovae: individual}

\maketitle

\section{Introduction}
\label{intro}

Ca-strong supernovae are a peculiar and poorly-understood class of transient, characterised most prominently by strong forbidden [Ca~II] emission in their nebular-phase spectra. For remote Ca-strong SNe (objects that occur at extended physical distances from their host galaxy), the prevalence of [Ca~II] emission has been interpreted as evidence for sub-Chandrasekhar mass white dwarf (WD) progenitors -- the ejecta of more massive progenitors would continue to burn into heavier elements than Ca, resulting in spectra that would no longer be dominated by [Ca~II] emission \citep{jacobson2020sn, polin2021nebular}. In addition to their strong late-time [Ca~II] emission, Ca-strong SNe share a distinct set of photometric and spectroscopic properties: low peak luminosities ($-$14.0 to $-$17.5 mag), rapid photometric evolution (rise times $\leq$ 16 days), intermediate photospheric velocities ($6000-11000$ km~s\textsuperscript{$-$1}), and fast transitions to the nebular phase \citep{kasliwal2012calcium}. Their tendency to appear remote from their host galaxies, in regions with no underlying star-formation, points to old and evolved progenitor systems \citep{lyman2014progenitors, lunnan2017two}. Despite their faint-and-fast evolving nature, and therefore low rate of discovery, volumetric rate estimates suggest that Ca-strong SNe may occur at $\geq15\%$ the rate of SNe Ia \citep{de2020zwicky}, with \cite{frohmaier2018volumetric} predicting they occur as frequently as 33-94$\%$. This makes them an important yet under-sampled population, with implications for both supernova physics and broader astrophysical processes, such as explaining the calcium enrichment observed in the intracluster medium \citep[ICM;][]{mulchaey2013calcium}.

In our previous study \citep[hereafter \citetalias{2025MNRAS.537.1015T}]{2025MNRAS.537.1015T}, we presented SN~2023xwi -- an otherwise typical Ca-strong SN, but notable for displaying strong [Ca~II] and [O~I] features in spectra taken around peak light. This was highly unusual: forbidden lines such as [Ca~II] are not expected to be present until much later phases. Under conventional explosion mechanisms, [Ca~II] emission in these objects is expected to occur \(\sim\)weeks after peak light when the ejecta have become optically thin and the photosphere has receded. This allows the central ejecta to evolve to a low enough density for forbidden emission to occur, and consequently ensures the SN is well into its nebular phase before this emission can be observed \citep{dessart2015one}. The presence of this line at such an early epoch raised the possibility that SN~2023xwi was an indicator of a broader, previously unrecognised behaviour in Ca-strong SNe.

A handful of Ca-strong SNe have been found to display an early photometric flux excess, or pre-peak bumps (PPBs), resulting in double-peaked light curves \citep[e.g.][]{jacobson2022circumstellar, ertini2023sn}. The physical mechanism responsible for these early flux excesses is yet to be fully understood. However, modelling of these early phenomena has shown they can be reproduced by shock cooling emission (SCE) or shock interaction with circumstellar material \citep{de2018iptf, jacobson2022circumstellar}. The second peak is assumed to be the result of radioactive \textsuperscript{56}Ni decay. Observations of SN~2019ehk revealed early X-ray emission; the origin of the early flux excess is not necessarily the same as that of the X-rays, but both phenomena indicate significant early-time emission in these objects \citep{jacobson2020sn, nakaoka2021calcium}. 

Several potential explosion models have been put forward to explain Ca-strong SNe, e.g., the Ca-strong object SN~2016hnk was shown to be consistent with both the thick He-shell double-detonation of a sub-Chandrasekhar mass WD \citep{polin2019observational, 2020ApJ...896..165J}, and the explosion of a near-Chandrasekhar-mass WD \citep{galbany2019evidence}. In \citetalias{2025MNRAS.537.1015T}, we proposed a progenitor scenario for the object SN~2023xwi: a recurrent He-nova AM Canum Venaticorum (AMCVn) system in which a white dwarf explodes into an H-poor environment polluted by prior nova eruptions.

In this paper, we investigate the observational behaviour of Ca-strong SNe using the largest sample to date. In Section~\ref{obs}, we outline our sample selection, along with their photometric, spectroscopic and host galaxy properties. In Section \ref{lighty time}, we present our analysis of the light curves, while in Section \ref{speccy time}, we present our spectroscopic analysis of the forbidden emission features, including investigation of transitional features between two distinct velocity components of the [Ca~II] emission line. In Section~\ref{lighty_ppb}, we investigate the early flux excess present in many Ca-strong objects, including modelling of the sample using the SCE model.  In Section~\ref{hi ca and o}, we discuss the implications of our results, and in Section~\ref{puppis_implicaysh} we discuss how the behaviours of Ca-strong SNe may be compatible with a polluted AMCVn progenitor scenario.

\section{Observations}
\label{obs}

In this section, we detail the sample selection and observations of the Ca-strong objects included in this analysis. In Section~\ref{ztf_sample}, we describe the sample selection of potential Ca-strong SNe from the Zwicky Transient Facility (ZTF). In Section \ref{new_photom}, we detail their photometric and spectroscopic observations, as well as those of one unpublished event from the Panoramic Survey Telescope and Rapid Response System \citep[Pan-STARRS; see][]{chambers2016pan}. In Section~\ref{sec:full_sample}, we discuss the additional literature Ca-strong SNe considered in our sample. 

\subsection{Sample selection criteria}
\label{ztf_sample}

ZTF is a wide-field optical time domain survey running on the 48-inch Schmidt telescope (P48) at the Mount Palomar Observatory \citep{2019PASP..131a8002B, 2019PASP..131g8001G}. The ZTF camera has a 47 square degree field of view and reaches \(\sim\) 20.5 mag in 30 s, allowing the survey to cover 3750 square degrees per hour at this depth \citep{2019PASP..131a8003M, 2020PASP..132c8001D}. ZTF's combination of depth and survey speed enable it to detect numerous transients, including Ca-strong SNe, at early times across a large fraction of the sky \citep{2019PASP..131g8001G}. As nebular spectroscopy is not achieved for all objects, and this is the primary diagnostic in determining a transient's status as a Ca-strong SN, we implemented similar criteria to \cite{de2020zwicky} to extract Ca-strong SNe from the full sample of ZTF objects between 2018 and 2024. Our primary sample within this time period was comprised of any objects spectroscopically classified as a Ca-strong SN using SNID \citep{2011ascl.soft07001B}, and any object that had been classified as SN Ib or Ca-Ia/Ib/Ic on the Fritz science platform\footnote{\url{https://fritz2024.caltech.edu/}}\citep{2019JOSS....4.1247V,2023ApJS..267...31C}, which implements the ZTF alert distribution system \citep{2019PASP..131a8001P}. Our selection criteria after this sample had been collated were:

\begin{enumerate}
    \item The SN must be fainter at peak than M\textsubscript{r}=$-$17.5 mag after correcting for Galactic reddening, assuming the same redshift as its host galaxy (or nearest viable host, see Section \ref{remoteness}). Host extinction is often negligible in these objects due to their typically remote locations. Ca-strong SNe with small (i.e. $\leq$3kpc) projected offsets from their host, however, are affected by host reddening. For the 8 objects in the full sample of candidate objects for which this was the case, we sourced the host extinction values from their associated publications, where available (see Table \ref{tab:full_sample}). For SN~2024jgw, due to incomplete photometric data, we could not infer the extinction. In instances where the photometric data were not complete enough to constrain the peak luminosity, we compared the decline of the light curve, along with any spectroscopic coverage, to the sample of confirmed Ca-strong objects. If the object did not deviate from the `expected' behaviour of Ca-strong objects and satisfied the following criteria, we assumed its peak luminosity fell within the expected range of M\textsubscript{r}>~$-$17.5 mag. For objects with no $r-$band photometric observations, we apply a peak \textit{g--r} colour correction of 0.51$\pm$0.07 mag (see Section \ref{t_peak}).
    \item The spectroscopic properties at peak should be consistent with hydrogen-poor SNe -- any objects with distinct broad hydrogen lines were excluded from the sample.
    \item The object must evolve into its nebular regime by +30d. This is defined as the combination of a fading continuum and the clear emergence of nebular emission lines, specifically [Ca~II]~\(\lambda\lambda\)7291,7324 and [O~I]~\(\lambda\lambda\)6300,6364 by this expected nebular-phase threshold. If the only data available are at much later epochs, \cite{de2020zwicky} include objects that display a complete transition to the nebular phase by +150 d. For objects with no spectroscopic coverage between +100 and +150~d, we have extended the cut-off for satisfying this criterion to the latest epoch within +200~d of peak light.
    \item As is standard, we require that any nebular-phase spectrum (after +30 d) that displays both forbidden [O~I]~\(\lambda\lambda\)6300,6364 and [Ca~II]~\(\lambda\lambda\)7291,7324 emission must have a line ratio strength of [Ca~II]/[O~I] > 2. If multiple spectra were available for an object, the [Ca~II]/[O~I] ratio was measured at the latest available spectroscopic epoch after +30~d.  We determined the lower limit of this ratio for objects with no significant [O~I] emission in their nebular-phase spectra by comparing the maximum possible integrated flux in the appropriate [O~I] emission region.
    \item The object must be spectroscopically distinct at early times from a SN Iax; if it has a strong photospheric-phase spectral match with an Iax on SNID, we reject it from the sample.
\end{enumerate}

\subsection{Newly classified Ca-strong SNe}
\label{new_photom}

Applying the selection criteria of Section \ref{ztf_sample} to the ZTF 2018-2024 sample, we identify 7 unpublished Ca-strong SNe that were ZTF follow-up targets -- SN~2021zfp, SN~2023er \citep[Gaia23aet, ATLAS23anl][]{2023TNSTR..57....1H}, SN~2023fhk, SN~2024uj \citep[ATLAS24api][]{2024TNSTR..85....1T}, SN~2024jgw, SN~2024uai, and SN~2024aaab. We also include SN~2023ugn, a PanSTARRS object, initially classified as a SN~Ib but that fulfils the Ca-strong selection criteria.

\begin{figure*}[!htbp]
	\includegraphics[width=\textwidth]{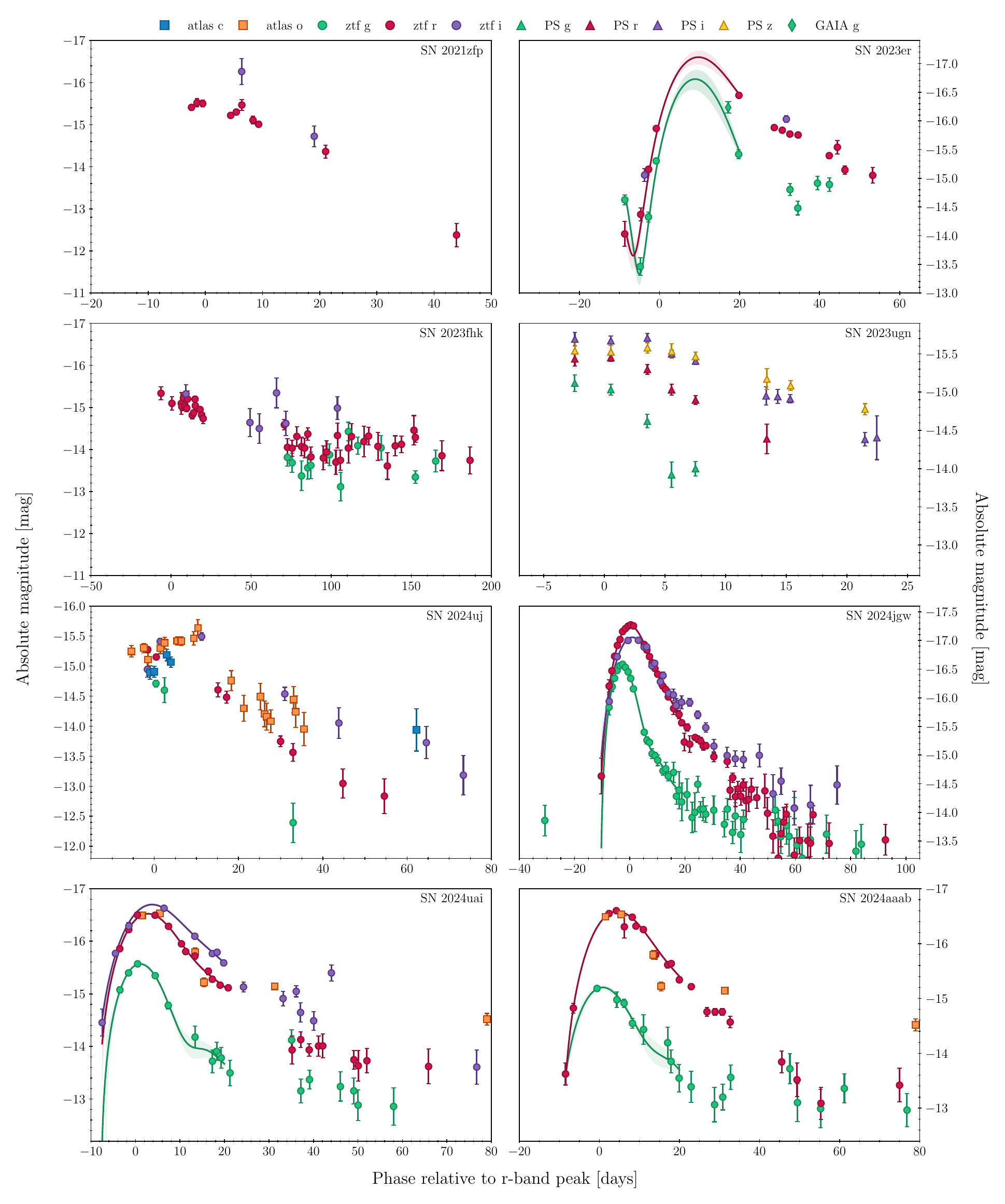}
    \caption{Optical light curves for each of the 8 confirmed new Ca-strong SNe in our sample. The optical data points are plotted with the following markers: ATLAS $c-$band -- blue squares, ATLAS $o-$band -- orange squares, ZTF $g-$band -- green circles, ZTF $r-$band -- red circles, ZTF $i-$band -- purple circles, PANSTARRS (PS) g-band -- green triangles, PS $r-$band -- red tringles, PS $i-$band -- purple triangles, PS $z-$band -- yellow triangles. Solid lines are the light curve fits as a result of our Gaussian processing fitting procedure (see Section~\ref{lighty time}).}
    \label{fig:new_lc}
\end{figure*}

Figure~\ref{fig:new_lc} shows the photometric evolution of these 8 newly-identified Ca-strong SNe. Seven of these objects were targetted by ZTF follow-up and were imaged in the ZTF $g,r,i$ bands, with supplementary data from the Asteroid Terrestrial-impact Last Alert System (ATLAS; \cite{tonry2018atlas, 2020PASP..132h5002S}) in $c$ (4100 $-$ 6600 \AA) and $o$ (5600 $-$ 8100 \AA) bands. The ZTF $gri$-band data was obtained using the \textsc{fpbot} package\footnote{\url{https://github.com/simeonreusch/fpbot}}; we removed data with unphysical flux errors, poor PSF fits, images taken through clouds, or those with a failure in the image processing \citep{rigault2025ztf}. SN~2023ugn was imaged in PS $g,r,i,z$ bands. 

We calculated the Milky Way (MW) extinction for each object using the \cite{2011ApJ...737..103S} recalibration of the \cite{schlegel1998maps} extinction maps, and corrected for this assuming a \cite{cardelli1989relationship} extinction law, with R\textsubscript{V} = 3.1. Key observational properties of each object (names, redshift, extinction correction, host galaxy name, projected offset from the host galaxy, and nebular [Ca~II]/[O~I] ratio (see Section \ref{sec:full_sample}), along with the associated references) are given in Table \ref{tab:full_sample}.

The photometric data enables constraints on peak magnitudes, times of peak light, colour evolution, and explosion epochs. Our methodology for determining these is discussed in Section~\ref{lighty time}. Of these 8 objects, only four objects were observed throughout the SN's rise, peak, and decline: SN~2023er, SN~2024gjw, SN~2024uai, and SN~2024aaab. A further three had data allowing the time of peak to be estimated through comparison with other Ca-strong objects: SN~2021zfp, SN~2023ugn, and SN~2024uj. The remaining object, SN~2023fhk was only observed during its decline so its time of peak light and peak $r-$band magnitude could not be constrained.

Spectra of the 8 newly identified Ca-strong SNe were sourced from a variety of telescopes. Each SN has between one and eight spectra at phases of $-7$ to +125~d, relative to inferred $r$-band peak (see Section \ref{t_peak}) -- these are presented in Figure~\ref{fig:new_spectra}.  The telescopes and instruments of the spectra are listed in Table \ref{tab:spec_info}, along with the phases relative to the time of $r$-band peak. We have corrected each spectrum for redshift and MW extinction, and have masked strong host galaxy emission features from the spectra of SN~2022psx and SN~2023fhk.

In each object with a spectrum after peak light (all except SN 2023er), there is a clear [Ca~II]~\(\lambda\lambda\)7291,7324 emission feature in the photospheric (\textless +14 d) spectra (often offset from rest wavelength by up to 6000 km~s\textsuperscript{$-$1}; see Figure~\ref{fig:all_vels_dist}), and often accompanied by the emergence of [O~I] within a few days. The spectroscopic evolution of the six SNe with multiple epochs is broadly consistent with previous observations of Ca-strong SNe, exhibiting relatively fast transitions from the photospheric to the nebular phase. While the signal-to-noise and wavelength coverage vary between SNe, the core features, particularly the [Ca~II] line, remain prominent in all events at later times. Unlike the other objects presented, SN~2023fhk grows redder as the SN evolves. This peculiar behaviour cannot be further investigated photometrically due to the lack of $g$-band observations at phases similar to those of the observed spectroscopy.

\subsection{The full Ca-strong sample}
\label{sec:full_sample}

In addition to these 8 new events, we collated published Ca-strong SNe and Ca-strong candidates, resulting in a total sample of 66 candidate Ca-strong SNe, at  $z$ of $\sim$0.004--0.04. Their key properties, along with their associated references are given in Table \ref{tab:full_sample}. These 66 events include 8 that were reclassified as Ca-strong SNe from other types (see Appendix \ref{reclassify} for further details). To ensure consistency, we applied the same cuts of Section~\ref{ztf_sample} to the full sample. Following criterion 4 of the selection criteria, we cut all candidates with [Ca~II]/[O~I]~<~2 after +30~d from peak light where data was available. For objects with no spectral observations after +30~d (marked by $\dagger$ in the [Ca~II]/[O~I] column of Table \ref{tab:full_sample}), we compared their available data with other Ca-strong SNe at similar phases. Both the photometric and spectroscopic properties of each object without nebular-phase spectral observations was consistent with the behaviours observed across the Ca-strong sample; we therefore continued to classify each as Ca-strong.

Applying these cuts removed 20 candidates. Two objects, SN~2019gsc and SN~2019ttf, were removed due to their prior classification of, and strong spectroscopic match with, SNe Iax. Of the 12 objects excluded due to a low [Ca~II]/[O~I] ratio, three (SN~2018jak, SN~2012hn, SN~2021pb) have previously been classified as Ca-strong \citep{2014MNRAS.437.1519V, 2023ApJ...959...12D}. However, as their measured [Ca~II]/[O~I] ratios fall below our adopted threshold (SN~2012hn: 1.63$\pm$0.01 at 149~d; SN~2018jak: 1.01$\pm$0.03 at 109~d; SN~2021pb: 1.27$\pm$0.01 at 109~d), we exclude them. SN~2018jak and SN~2021pb additionally exhibit broad H lines in their photospheric-phase spectra, which independently disqualifies them based on our selection criteria. The remaining six objects were disqualified for displaying broad H lines in their photospheric spectra. Our final sample consists of 46 confirmed Ca-strong SNe.

\subsection{Host galaxy properties}
\label{remoteness}

For objects with no previously assigned host galaxy, we determined the host to be the nearest galaxy with a measured redshift that was consistent with the forbidden [Ca~II] NIR triplet emission line in the latest spectral epoch for that given object. All objects in the sample of Ca-strong objects had a viable candidate host with a measured spectroscopic redshift; we therefore assumed the redshift of each object to be that of its host galaxy, and calculated the luminosity distances using these redshift values, correcting for host-galaxy peculiar velocities where available. We determined the projected physical offset of each object from its host by measuring the angular separation between the transient position and the host galaxy centroid and converting this to a projected physical distance (Table \ref{tab:full_sample}). 

%I'm commenting this out because it seems similar to previous sentences. Due to the historically small sample size of Ca-strong objects, the behaviours of a few objects have become the expected standard across the whole class. Objects such as iPTF11kmb -- one of the most prototypical Ca-strong objects, and often a benchmark object when classifying other Ca-strong candidates -- and SN~2023er have substantial offsets of >100 kpc from their hosts. This has resulted in a skewed perception of the typical separation between Ca-strong SNe and their hosts.

\begin{figure}%[!htbp] Plots should go to top of page
	\includegraphics[width=\columnwidth]{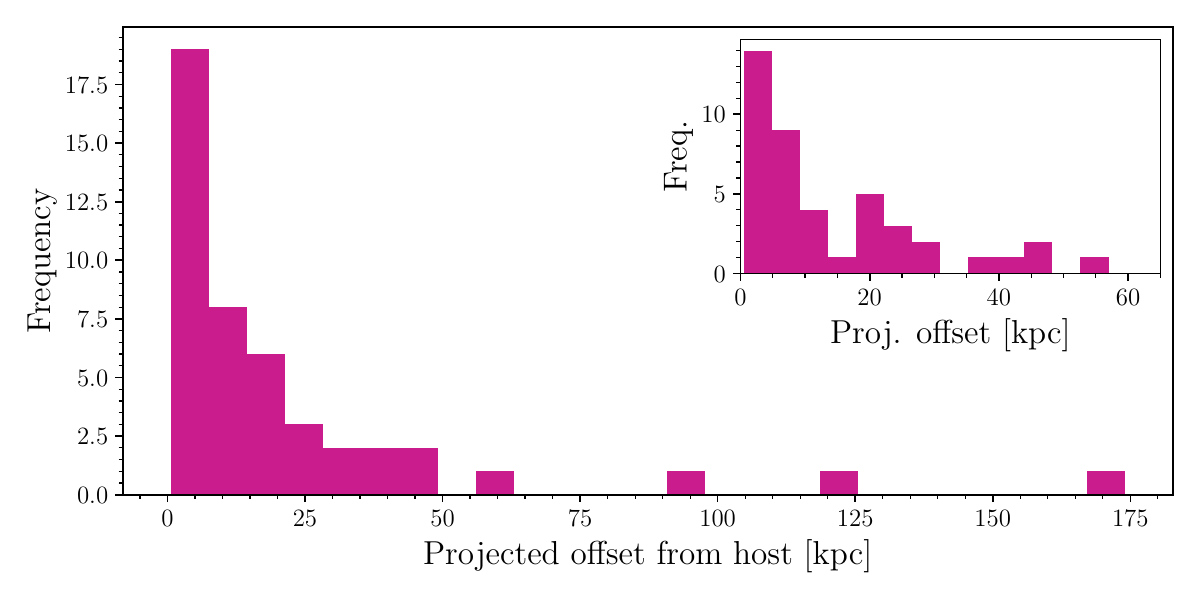}
    \caption{Histogram of the physical separation of each Ca-strong object in our sample from its host. Inset: Histogram of objects with separations of $<$60 kpc, removing three objects (iPTF11kmb, SN~2019bkc, SN~2023er).}
    \label{fig:remoteness}
\end{figure}

As discussed in Section \ref{intro}, Ca-strong transients display a preference for early-type galaxies, as well as group and cluster environments, suggestive of old stellar populations \citep{lyman2014progenitors, lunnan2017two}. Figure~\ref{fig:remoteness} shows a histogram of the separation of confirmed Ca-strong objects in our sample from their hosts. The weighted mean of the distribution $5.76\pm0.75$ kpc, and a median of $9.90$ kpc. \cite{lunnan2017two} found a median projected offset of $8 - 30$ kpc (dependent on the exact sample definition) for a previous smaller literature sample, while \cite{Srivastav2026} using a sample of Ca-strong transients discovered by ATLAS (and which are a subset of our sample) found a median projected offset of $\sim$11 kpc.  Our median value is slightly lower than these but not significantly different. As shown by previous samples such as \cite{lunnan2017two}, the projected offset distribution of Ca-strong transients is more offset than the normal SN Ia distribution and \(\sim\) 25\% of the objects in our sample still occur at offsets of > 20 kpc (and two at >100 kpc). These trends imply, therefore, that these objects can still be considered `remote' objects compared to typical SNe Ia.

\section{Photometric analysis}
\label{lighty time}

For objects with well-sampled light curves around peak, we fitted them using a Gaussian Process (GP) regression (see Figure~\ref{fig:new_lc}). GP regression is a non-parametric Bayesian approach that models the light curve as a distribution over functions, rather than assuming a fixed template \citep{williams2006gaussian}. This makes it particularly well-suited for our sample of SNe, wherein cadence and noise properties vary between objects \citep[e.g.][]{2025A&A...694A..10D}. We adopted a squared exponential kernel with an added white-noise term to describe the smooth evolution of the light curves in combination with associated photometric uncertainties and varied-cadence data for each object in our sample \citep{roberts2013gaussian}. We optimised the length scale and amplitude of the kernel for each object using \texttt{emcee} to determine the parameters with the maximum likelihood.

\subsection{Peak light properties}
\label{t_peak}

From the GP fits, we determined the time of peak brightness in order of preference: $r$, $g$, $i$, and finally ATLAS $c$ and $o$ bands. Since the time of peak can differ between filters, prioritising $r$-band ensured consistency across the sample given its most consistent coverage. Uncertainties on the time of peak $r$-band brightness, $t_{\rm peak}$, were estimated from the GP predictive variance, marginalised over the kernel parameters. Additionally, the GP fits provided measurements of the absolute magnitude at peak, $g-r$ colour at peak, and the $r$-band decline rate, $\Delta m_{7}(r)$, defined as the change in magnitude in the first 7 days post-maximum light (Table \ref{tab:photometric_params}). 

Of the full sample of 46 confirmed Ca-strong SNe, 34 objects had sufficient optical light curve data for determining $t_{\rm peak}$, including four of the new SNe shown in Figure~\ref{fig:new_lc}. For the 8 literature objects with no observations in the appropriate range (SN~2000ds, SN~2001co, SN~2003H, SN~2003dg, SN~2005E, iPTF15eqv, SN~2018lqu, SN~2018gwo), we adopted times of $r$-band maximum from their discovery papers. For the new objects with no photometry around peak light (SN~2021zfp, SN~2023fhk, SN~2023ugn, SN~2024uj), we estimated their time of peak by cross-matching their photometry with more well-constrained Ca-strong objects in our sample. As SN~2023fhk was discovered post-peak, the first detection of this object at MJD~60036 marks the upper limit of its time of peak light. Where $r$-band data were not available, we determined photometric behaviour from $g$-band data, applying a correction of $t_{\rm peak_{r}}=t_{\rm peak_{g}}+2$~d, based on the mean \(\delta\)t\textsubscript{r,peak-g,peak} of 2.0\(\pm\)0.2~d across the sample of objects with sufficient coverage in both bands around peak light.

\begin{figure}
	\includegraphics[width=\columnwidth]{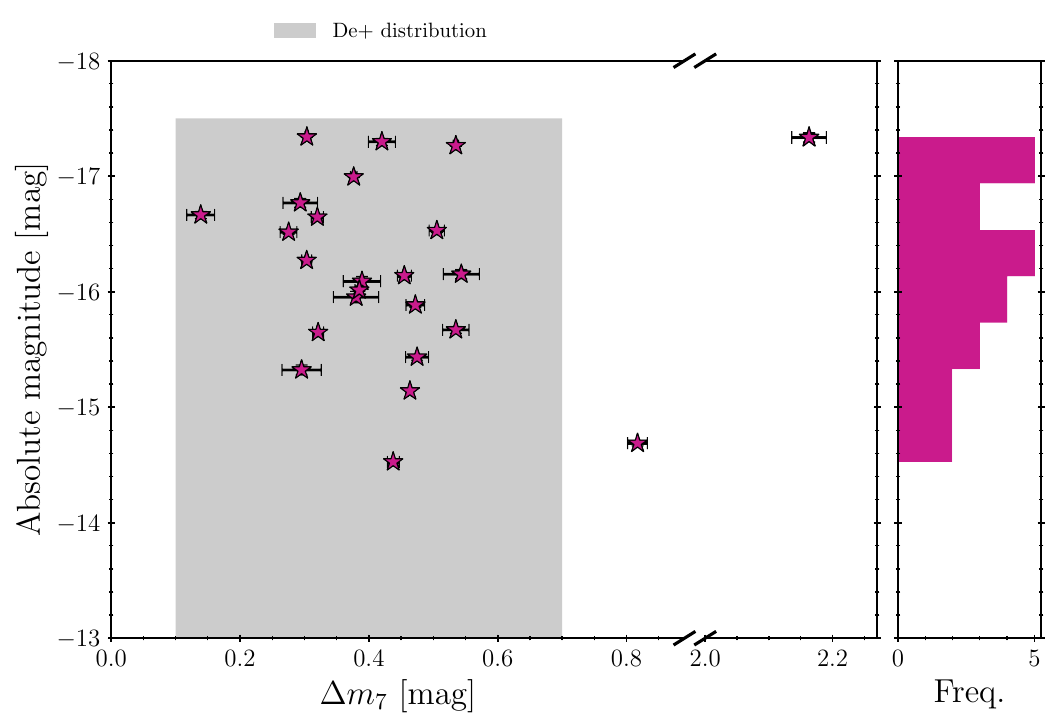}
    \caption{Absolute peak $r-$band magnitude against $\Delta m_{7}(r)$. The parameter space for Ca-strong SNe from \citet{de2020zwicky} is shaded in grey.}
    \label{fig:decline_rate}
\end{figure}

Figure~\ref{fig:decline_rate} shows the distribution of absolute peak $r-$band magnitudes and $\Delta m_{7}(r)$ -- the decline in $r-$band magnitude in 7 days following peak light -- for the 24 objects of the full Ca-strong sample with $r-$band observations around peak light. Consistent with the previous study of \cite{de2020zwicky}, which is shown in grey in Figure~\ref{fig:decline_rate}, the population is confined to faint peak luminosities ($M_r \gtrsim -17.5$ mag) and rapid post-peak declines ($\Delta m_{7}\geq 0.1$ mag). All objects bar two in our sample -- SN~2021inl and SN~2019bkc \citep[which has been independently noted for its extraordinarily rapid evolution, see][]{prentice2020rise}, which evolve faster than the rest of the sample -- fall within this expected parameter space, reinforcing the classification of Ca-strong SNe as faint, fast-evolving explosions. We note, however, that the definition of this class is partially circular, since the faintness (peak $r$-band magnitude fainter than $-$17.5 mag) and rapid evolution are themselves used in the classification. To ensure that this faintness requirement did not introduce bias by excluding more luminous, yet otherwise `typical', Ca-strong SNe, we examined all objects in our original sample satisfying the remaining classification criteria but failing the brightness cut. No such objects were identified.

We determined the \textit{g--r} colour at peak of the 24 Ca-strong objects with peak photometry in both $g-$ and $r-$bands. All are red at peak; the distribution is peaked at a \textit{g--r} colour of 0.51$\pm$0.07 mag, with colour ranging from 0.42$\pm$0.01 to 1.64$\pm$0.01 mag. This exceptionally red object, which is an outlier from the rest of our sample, is SN~2021sjt. This SN is located in a spiral arm, suggesting this may be at least partially explained by host-galaxy extinction.

\section{Spectroscopic analysis of forbidden emission features}
\label{speccy time}

\citetalias{2025MNRAS.537.1015T} found that forbidden transition emission lines,  specifically [Ca~II]~\(\lambda\lambda\)7291,7324, [O~I]~\(\lambda\lambda\)6300,6364, and [O~I]~\(\lambda\)5577, can be observed from as early as peak light. This observation, which is in contradiction with traditional understanding (i.e. that these nebular emission lines should only be visible when the ejecta have expanded and cooled) is a potential indicator of a more complex progenitor system than previously thought. \cite{2025A&A...702A..29C} find the merger of a $0.6~\textrm{M}_{\odot}$ CO WD and $0.4~\textrm{M}_{\odot}$ He WD can reproduce spectroscopic evolution consistent with Ca-strong SNe. In their model, [Ca~II] emission begins to become prevalent at +20~d -- earlier than the typical threshold of +30~d by which Ca-strong SNe are expected to evolve to their nebular regime \citep{de2020zwicky}. This emission, however, is still substantially delayed from the observed peak light [Ca~II] emission of SN~2023xwi presented in \citetalias{2025MNRAS.537.1015T}. To quantify the extent of this phenomenon -- i.e. whether the early emission of these forbidden features was unique to this object, or if it is common across the sample of Ca-strong SNe -- we investigated its presence and behaviour throughout the sample of Ca-strong objects. 

\subsection{Single Gaussian fitting to forbidden emission lines}
\label{vel measure}

The non local thermodynamic equilibrium (NLTE) model of a He-shell detonation of a CO WD presented by \cite{dessart2015one} produces faint and fast light curves, similar to those of Ca-strong SNe. Their synthesised spectra produce strong Ti~II features in the spectral region we identify as [Ca~II] emission, and blended Sc~II + Ti~II contributions in the [O~I] emission region. These models are significantly reddened due to line-blanketing as a result of the large Ti~II abundance -- a phenomenon we do not observe in Ca-strong SNe. Similarly, the model presented by \cite{2025A&A...702A..29C} is significantly reddened due to the predicted high abundance of Ti~II. We do not observe this high degree of reddening across the sample of Ca-strong SNe at peak light, and therefore assume the features in these regions to be [Ca~II] and [O~I], respectively.

\begin{figure}
	\includegraphics[width=\columnwidth]{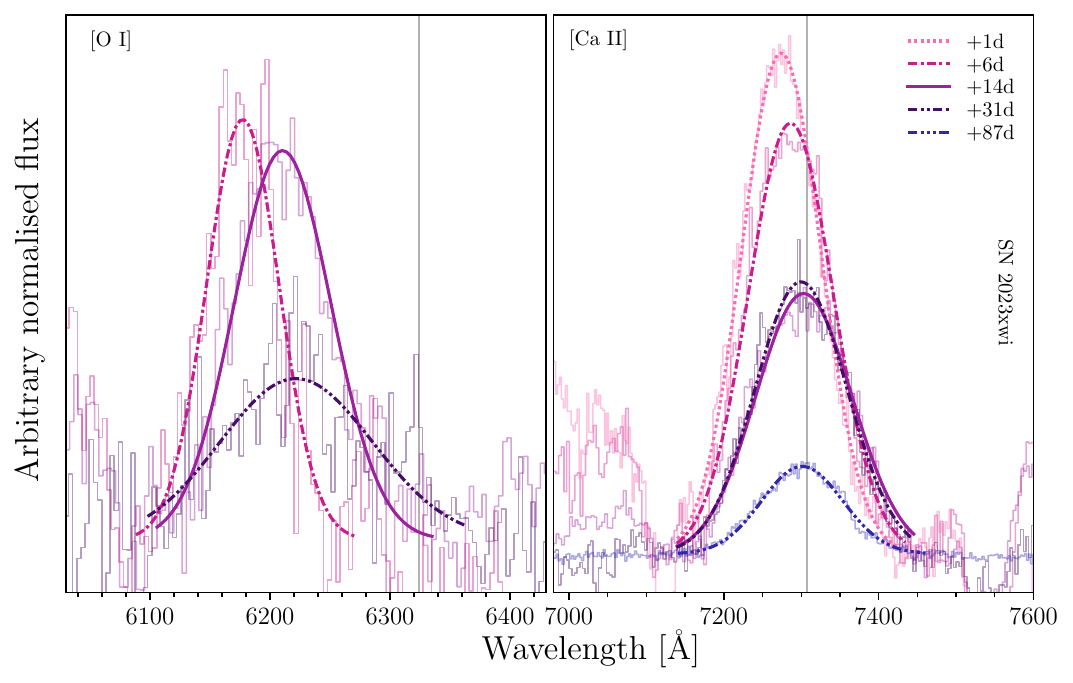}
    \caption{Example local pseudo-continuum-corrected emission line model fit of the [O~I] (left) and [Ca~II] (right) feature in SN~2023xwi throughout its evolution. The spectra are overlaid with their total model profile fit. The vertical grey line shows the weighted average of the two component wavelengths of each forbidden emission feature at rest.}
    \label{fig:example_model_fit}
\end{figure}

To model the emission features of each of the forbidden emission lines of [Ca~II]~\(\lambda\lambda\)7291,7324, [O~I]~\(\lambda\lambda\)6300,6364, we modelled each feature as a combination of the local pseudo-continuum and the profile of the emission lines. The local pseudo-continuum was approximated by a linear relation between wavelength and observed flux between the two local minima of the feature. The emission profile of both the [Ca~II] and [O~I] features were modelled with a singular Gaussian profile as the two component emission lines of these doublets could not be confidently resolved in any spectra. Due to convolution with the instrument resolution of each spectrum \citep{2012MNRAS.420.3451M}, a single Gaussian model is the simplest plausible empirical analogue to the observed emission profiles. We fit all spectra for every object in our sample using this methodology. Example model fits to the [Ca~II] and [O~I] regions for SN~2023xwi are shown in Figure~\ref{fig:example_model_fit}. 

Uncertainties were calculated using a Monte Carlo approach, in which the start and end points of the fitting region were varied based on the resolution of any given spectrum. For higher-resolution spectra (resolution < 2 $\AA$), the fitting boundaries were perturbed by nine times the spectral resolution. Medium-resolution spectra (2 – 10 $\AA$) were varied by six times the resolution, while low-resolution spectra (resolution > 10 $\AA$) were permitted a maximum fixed variation of 55 $\AA$. Each model was refitted 1000 times per spectrum, with random sampling of the start and end points; the mean and standard deviation of the resulting velocities were taken as the measured velocity and its associated uncertainty, respectively. Fits in which the uncertainty was both greater than 300 km~s\textsuperscript{$-$1} and exceeded 20\% of the mean value of the measured velocity offset were further investigated to ensure the measured velocity was associated with the correct spectral feature. This uncertainty was then combined with the uncertainty associated with the measured redshift, as well as the uncertainty associated with the resolution of the spectrum itself, resulting in the total velocity uncertainty at each spectral epoch.

In cases where both the [O~I]~\(\lambda\lambda\)6300,6364 and \(\lambda\)5577 features were present, the velocity of the [O~I]~\(\lambda\)5577 line was used as a check to verify the identification of the [O~I]~\(\lambda\lambda\)6300,6364 feature -- the velocity offset associated with each of these two [O~I] emission lines has been shown to be in agreement within 3\(\sigma\) throughout an object's full spectral evolution (see Section~5.2 of \citetalias{2025MNRAS.537.1015T}). If the measured [O~I]~\(\lambda\lambda\)6300,6364 and ~\(\lambda\)5577 velocities agreed within a 3\(\sigma\) significance level, the identification of the feature as [O~I] was deemed reliable. For objects with no clearly distinguishable [O~I]~\(\lambda\)5577 emission, if the measured velocity of the presumed [O~I]~\(\lambda\lambda\)6300,6364 feature was within the range of expected values ($-$14000 to +2500 km~s\textsuperscript{$-$1}, derived from spectra with confirmed [O~I] emission) we assumed the feature had been correctly identified as [O~I]. 

To ensure the same feature was being measured at all epochs for a given SN, the pseudo-equivalent widths of each feature were measured at each epoch. A feature was confirmed to be the same spectral emission line for pseudo-equivalent widths that were consistent across all photospheric epochs within a 5\(\sigma\) threshold across all observations before +14~d of peak light. In the photospheric phase, the continua of Ca-strong SNe is complex and therefore, the 5\(\sigma\) limit was chosen to account for any discrepancies in the pseudo-continuum approximation. For all analysed lines in our sample, no such features were re-classified.

In our full sample of 46 Ca-strong objects, all displayed prominent [Ca~II] emission. 42 objects had distinct [O~I]~\(\lambda\lambda\)6300,6364, 35 of which were verified by their [O~I]~\(\lambda\)5577 feature. Of the 35 objects in our full sample which were spectroscopically observed within 7~d of peak light, 29 had measurable forbidden [Ca~II] emission in their peak spectra. If this is extended to within 10~d of maximum, 32 of the 35 objects with spectra around this epoch display distinct [Ca~II] emission. In all objects with later observations, this feature persisted in the spectra through the nebular phase, remaining visible in the latest spectrum of each object.

\begin{figure*}[!htbp]
	\includegraphics[width=\textwidth]{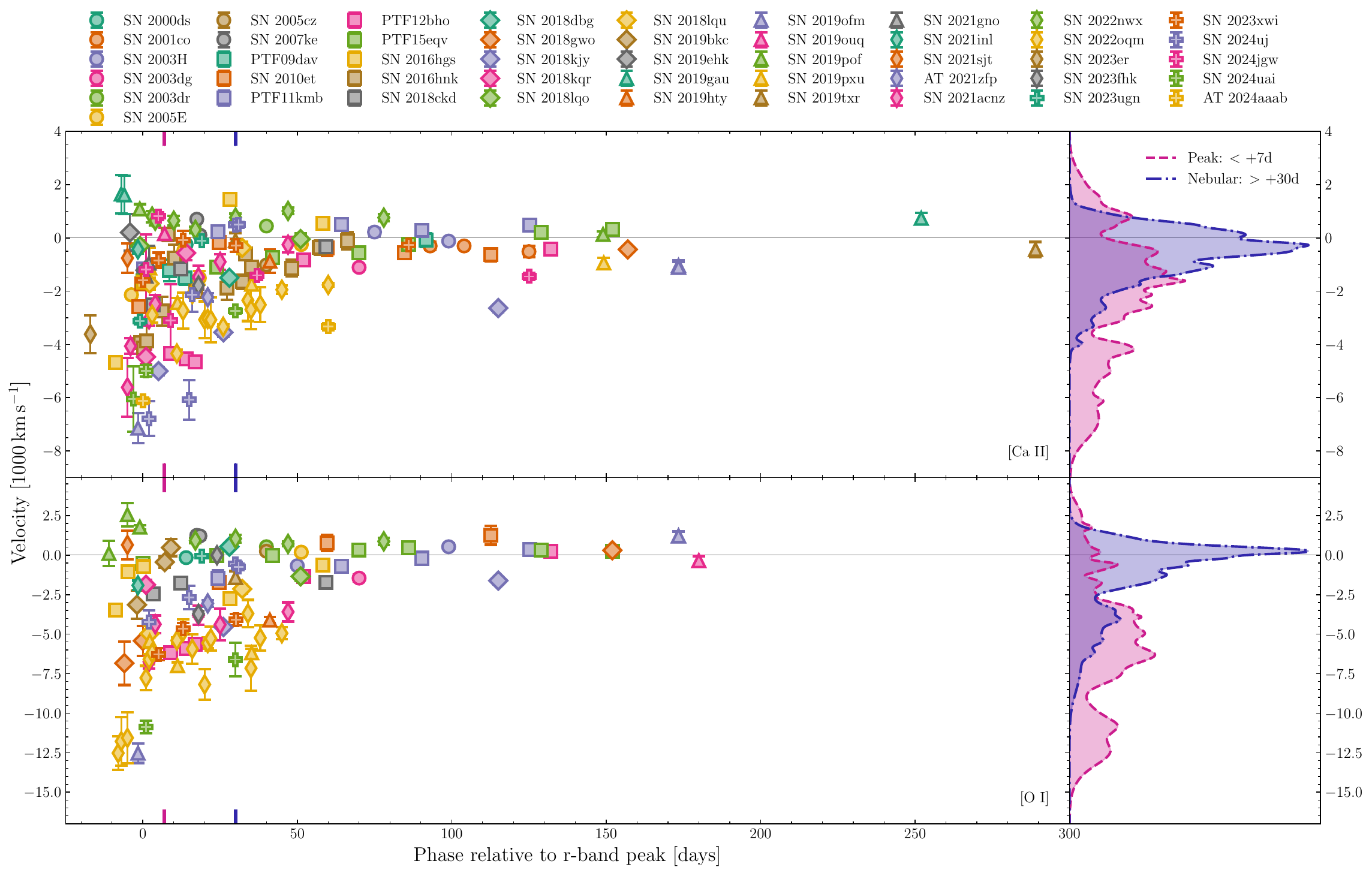}
    \caption{Left panels: Measured forbidden emission line velocities of [Ca~II]~\(\lambda\lambda\)7291,7324 (upper) and [O~I]~\(\lambda\lambda\)6300,6364 (lower). Right panels: Distribution of velocities for the forbidden transition emission [Ca~II] (upper) and [O~I] (lower). The peak (pink dashed, spectrum closest to peak per SN at < +7d) and nebular (blue dot-dashed, latest spectrum taken per SN at > +30d) distributions -- constructed by summing the Gaussian probability distributions of each velocity -- are overlaid and normalised with respect to the nebular distributions. The cut-off point for the peak distribution (+7d) and the start of the nebular distribution (+30d) are shown by pink and blue tick marks, respectively, on the left subplots.}
    \label{fig:all_vels_dist}
\end{figure*}

\subsubsection{Distribution of forbidden emission line velocities}
\label{vel dist}

The velocity evolution of the [Ca~II] and [O~I]~\(\lambda\lambda\)6300,6364 features (relative to their mean rest-frame wavelengths) for each SN are shown in the upper and lower panels of Figure~\ref{fig:all_vels_dist}, respectively. 

The velocities of both forbidden emission features generally evolve from a broad range of non-zero velocities in the photospheric spectra -- defined as those before +30~d from peak -- to a velocity consistent with zero velocity at late times. The pink dashed histograms in the right panels of Figure~\ref{fig:all_vels_dist} show the distribution of velocities from spectra taken closest to peak brightness for each object, all within +7~d of peak. 
 
We measured the weighted mean velocities of the distribution of peak spectra with respect to their associated uncertainties to be $-$2290$\pm$830 km~s\textsuperscript{$-$1} for [Ca~II] and $-$6040$\pm$1400 km~s\textsuperscript{$-$1} for [O~I]~\(\lambda\lambda\)6300,6364.  The distribution of [Ca~II] velocity has a broad spread of \(\sigma=\pm\)2260 km~s\textsuperscript{$-$1}, and extreme red-shifted and blue-shifted velocities spanning 1600$\pm$700 km~s\textsuperscript{$-$1} (SN~2019gau at $-$7~d) and $-$6800 $\pm$700 km~s\textsuperscript{$-$1} (SN~2024uj, +2 d), respectively. Similarly, the peak velocity distribution of [O~I] is broad with \(\sigma=\pm\) 4110 km~s\textsuperscript{$-$1}, and extremes of 2600$\pm$700 km~s\textsuperscript{$-$1} (SN~2019pof, $-$5 d) and $-$12500$\pm$1000 km~s\textsuperscript{$-$1} \citep[SN~2022oqm, $-$8 d;][]{2024ApJ...962..109I}. Any objects with a red-shifted velocity around peak light (as opposed to a blue-shifted velocity as seen across the rest of the sample) have emission line offsets consistent with 0 km~s\textsuperscript{$-$1} to, at most, a level of 3\(\sigma\).

In contrast, the nebular-phase spectral velocities, shown as blue dot-dashed histograms in Figure~\ref{fig:all_vels_dist}, corresponding to the latest available spectra for each SN (taken after +30~d from peak) exhibit notably narrower distributions than at maximum light. The mean velocities of these late-time spectra are $-690\pm350$ km~s\textsuperscript{$-$1} for [Ca~II] and $-1550\pm810$ km~s\textsuperscript{$-$1} for [O~I]~\(\lambda\lambda\)6300,6364, which are consistent with zero velocity within 2\(\sigma\).

The transition from a broad and offset distribution at peak, to a narrower, near-zero distribution at late times may be the result of emission originating from different regions of Ca-strong SNe as they evolve. The forbidden emission we observe in the nebular-phase spectra is likely a result of core-dominated emission as expected at these epochs. Conversely, the presence of these forbidden transition emission lines around peak light, but with distinctly non-zero velocities, implies a separate origin for the early-time emission. This early-phase emission may arise from material that is pre-existing and external to the SN explosion itself, as observed in SN~2023wxi (\citetalias{2025MNRAS.537.1015T}). The physical interpretation is discussed further in Section~\ref{vel_ev_dist}.

In addition to the evolution of each forbidden line, throughout the full spectral evolution of each object in the sample of Ca-strong SNe, only 5 spectra show [O~I] emission associated with a `slower' intrinsic velocity than [Ca~II] at the same epoch with a significance level of $>3\sigma$ -- in all other cases, the [O~I] emission is blueshifted to a greater velocity than the [Ca~II].

\subsection{Two-component [Ca~II] fitting}
\label{vel double fit}

\begin{figure*}
    \centering
    \includegraphics[width=0.47\textwidth]{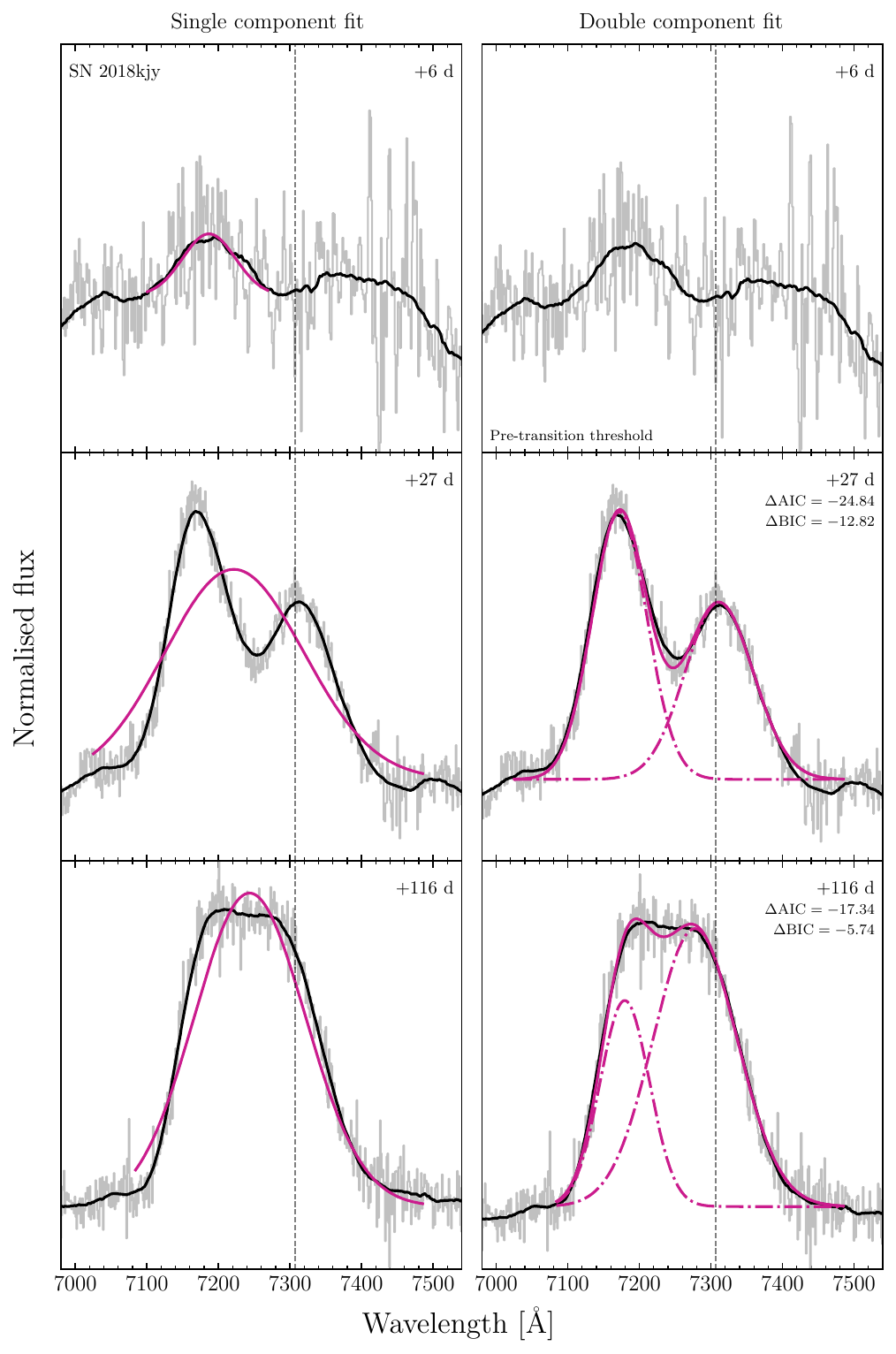}
    \hfill
    \includegraphics[width=0.47\textwidth]{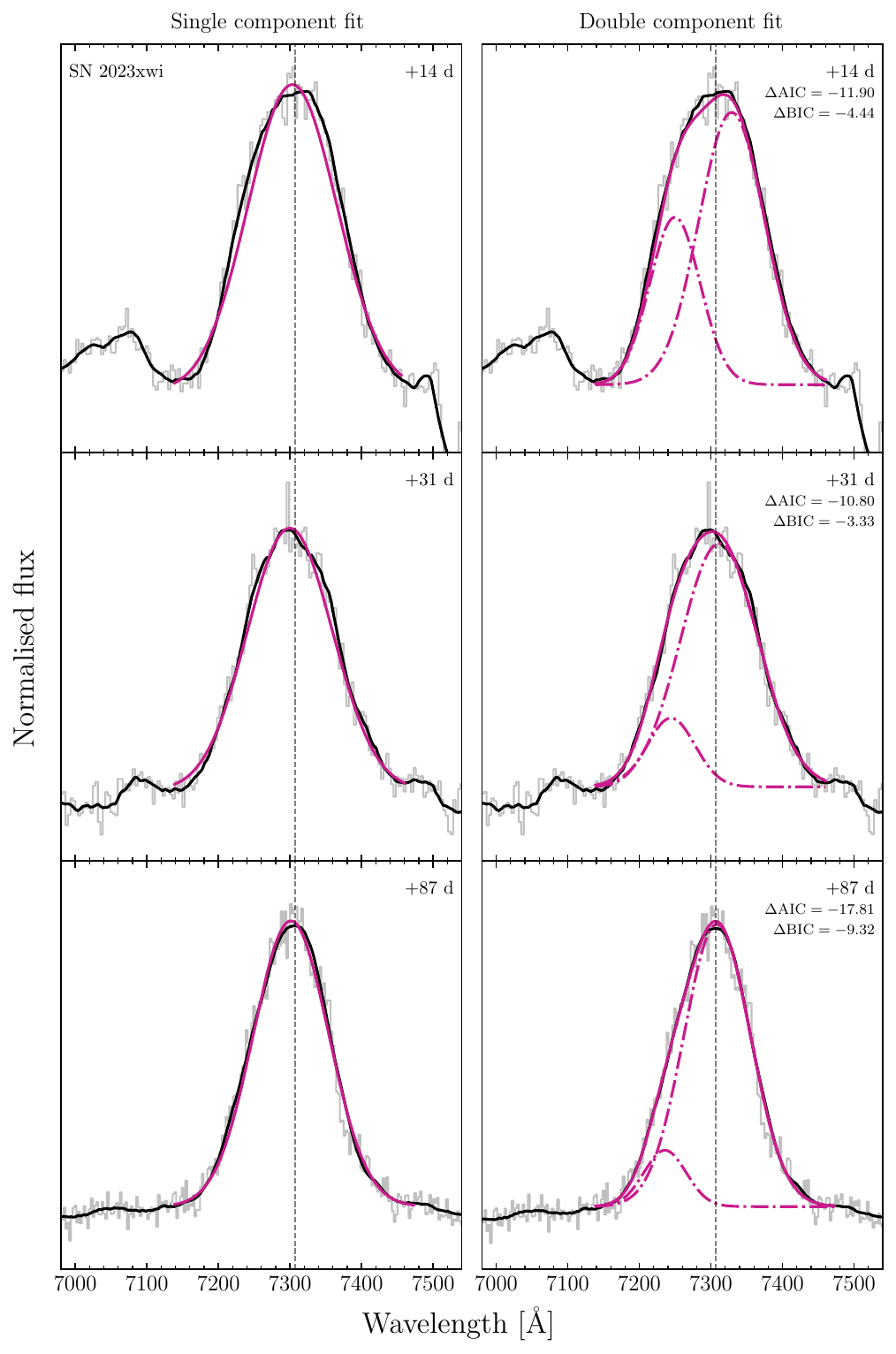}
    \caption{Example [Ca~II] single- and double-component Gaussian mixture model fits for SN~2018kjy (left) and SN~2023xwi (right). The pseudo-continuum-corrected spectrum (light grey) is overlaid with a Savitzky-Golay smoothed spectrum for display purposes (black). The total model profile fit is shown with a solid pink line. Dot-dashed pink lines show the component Gaussians. The left panel for each object shows the single Gaussian fit across the full spectral time series, and right panel shows the double Gaussian profile for each of the spectra observed after the transition threshold. The vertical dashed line corresponds to the weighted mean rest wavelength of the [Ca~II] feature.}
    \label{fig:example_model_fit_x2}
\end{figure*}

\citetalias{2025MNRAS.537.1015T} hypothesised the presence of `transition features' in the spectra of Ca-strong SNe. These are the spectral signatures of the transitional phase between the early-time offset velocity feature and core-dominated emission feature of the SN. Of the sample of 46 Ca-strong SNe, we have spectra at a high enough resolution in the correct phase range (\(\sim\) +14 -- +30~d from peak) for 26 objects to study these transition features. At earlier phases, the traditionally `nebular' emission is obscured by the photosphere, and we did not model this spectral feature at early times with the two-component model. Due to contaminating emission lines at the wavelength of the [O~I] feature, as well as its weaker strength into the nebular phase compared to [Ca~II], our investigation into transition features is focussed solely on the [Ca~II] feature.

\begin{figure*}
	\includegraphics[width=\textwidth]{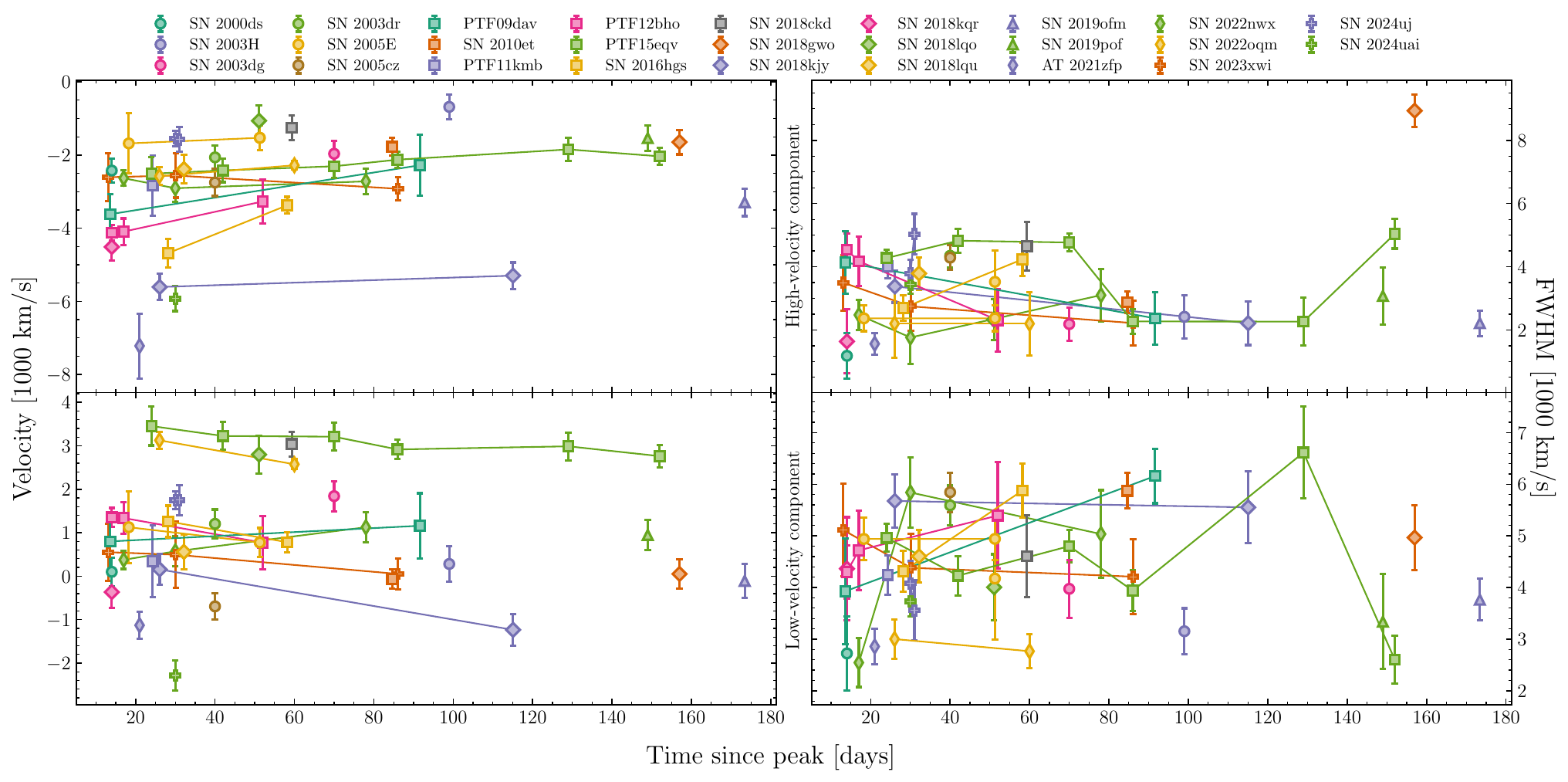}
    \caption{Velocity and FWHM evolution of the double-component Gaussian model for the [Ca~II] feature. The left sub-panel shows the velocity evolution of the high-velocity component (upper) and low-velocity component (lower). The right sub-panel shows the FWHM evolution of the high-velocity component (upper) and low-velocity component (lower).}
    \label{fig:ca_x2}
\end{figure*}

To determine if transitional features were present, the [Ca~II] emission profile in spectra after +14d was modelled as a model comprised of two Gaussian profiles (one for the higher-velocity offset emission and one for the lower-velocity emission) after we continuum-corrected each spectrum. Uncertainties were calculated using the same Monte Carlo-based approach as outlined in Section~\ref{vel measure}. The inclusion of an additional Gaussian component introduces two degrees of freedom to the model, thus will often produce a lower residual than the single-Gaussian model. We therefore evaluate each model using criteria which penalise this increased model complexity: the Akaike Information Criterion \citep[AIC,][]{akaike1974new} and Bayesian Information Criterion \citep[BIC,][]{stone1979comments}. They are defined as:
\begin{equation}
   \textrm{AIC} = -2 \ln{(\textrm{L})} + 2k
\end{equation}
and
\begin{equation}
   \textrm{BIC} = -2 \ln{(\textrm{L})} + k \ln{(\textrm{N})} \textrm{,}
\end{equation}
with likelihood, \textit{L}, number of model parameters, \textit{k}, and number of data points, \textit{N}. For each best-fit, we calculated the AIC and BIC values using \texttt{RegscorePy} -- a Python package that aids model comparison  with the use of various regression models. AIC does not penalise as harshly as BIC, and therefore the favoured model using AIC may be over-fitting the data. Conversely, BIC penalises harshly for additional parameters, and will favour the simplest appropriate model. This can result in a favoured model that is an over-simplification of the underlying physical mechanisms. Due to these intricacies, it is best to optimise a fit with respect to both the AIC and BIC values. 

Negative values of \(\Delta\)AIC and \(\Delta\)BIC (where \(\Delta\)AIC(BIC) is the difference between the AIC(BIC) values for the double- and single-component Gaussian models per epoch) indicate a preference for the double-component model, with values of $\Delta \rm{AIC}\leq-5$ and $\Delta \rm{BIC}\leq-2$ indicating a strong preference for the two-component model of [Ca~II] emission during the transitional phase \citep{kass1995bayes, anderson2004model}. These values indicate that the fit improvement exceeds that expected solely from the additional free parameters in the double-component model. In Table \ref{tab:aic_bic}, we present \(\Delta\)AIC and \(\Delta\)BIC between double- and single-component Gaussian models for each spectrum in the transitionary period, as well as later time spectra to determine if these features persist.

All Ca-strong objects with spectra within the expected phase range of transitional features (~+14 to +30~d) prefer the double-Gaussian fit, indicative of two distinct emission components, potentially arising from two distinct emitting regions. Example single- and double-component fits to the [Ca~II] feature are shown for SN~2018kjy in the left panels and for SN~2023wxi in the right panels of Figure~\ref{fig:example_model_fit_x2}.  This preference for a double-peaked emission profile is not due to the doublet nature of the [Ca~II] emission feature as the velocity separation of our measured two components is significantly larger than the separation of the [Ca~II] doublet component lines in all cases. 

While the double-component Gaussian model is preferred throughout the sample of Ca-strong SNe, we note this does not necessarily require two distinct emitting regions; alternative ejecta structures and geometries can produce emission line profiles that are similarly approximated by a double-component Gaussian model. A shell-like distribution of Ca, for example, could produce a flat-topped [Ca II] profile, while a disc or ring-like structure could produce a `double-horned' emission line \citep{2017hsn..book..795J}. As we observe the emergence of a persistent `high-velocity' component at a significantly earlier epoch than the `low-velocity' component, we favour a system in which the two Gaussian components are representative of two distinct emitting regions. In such a system, the low-velocity component arises from a secondary emitting region as the SN transitions into the nebular regime. While we observe this trend in Ca-rich SNe with multiple spectral observations from peak light, through the transitional phase, and into the truly nebular regime, we cannot definitively conclude that this behaviour is consistent throughout the sample of Ca-strong SNe. The observed velocity evolution associated with the forbidden [Ca~II] and [O~I] emission lines is consistent with our interpretation of two distinct emitting regions, and we therefore favour this interpretation. We note, however, that as multi-epoch spectra spanning the relevant evolutionary phases were not observed for each object, we cannot definitively rule out the possibility that Ca-strong SNe constitute a heterogeneous class in which similar observed profiles and evolution arise from different underlying ejecta geometries.

\begin{figure}[!htbp]
	\includegraphics[width=\columnwidth]{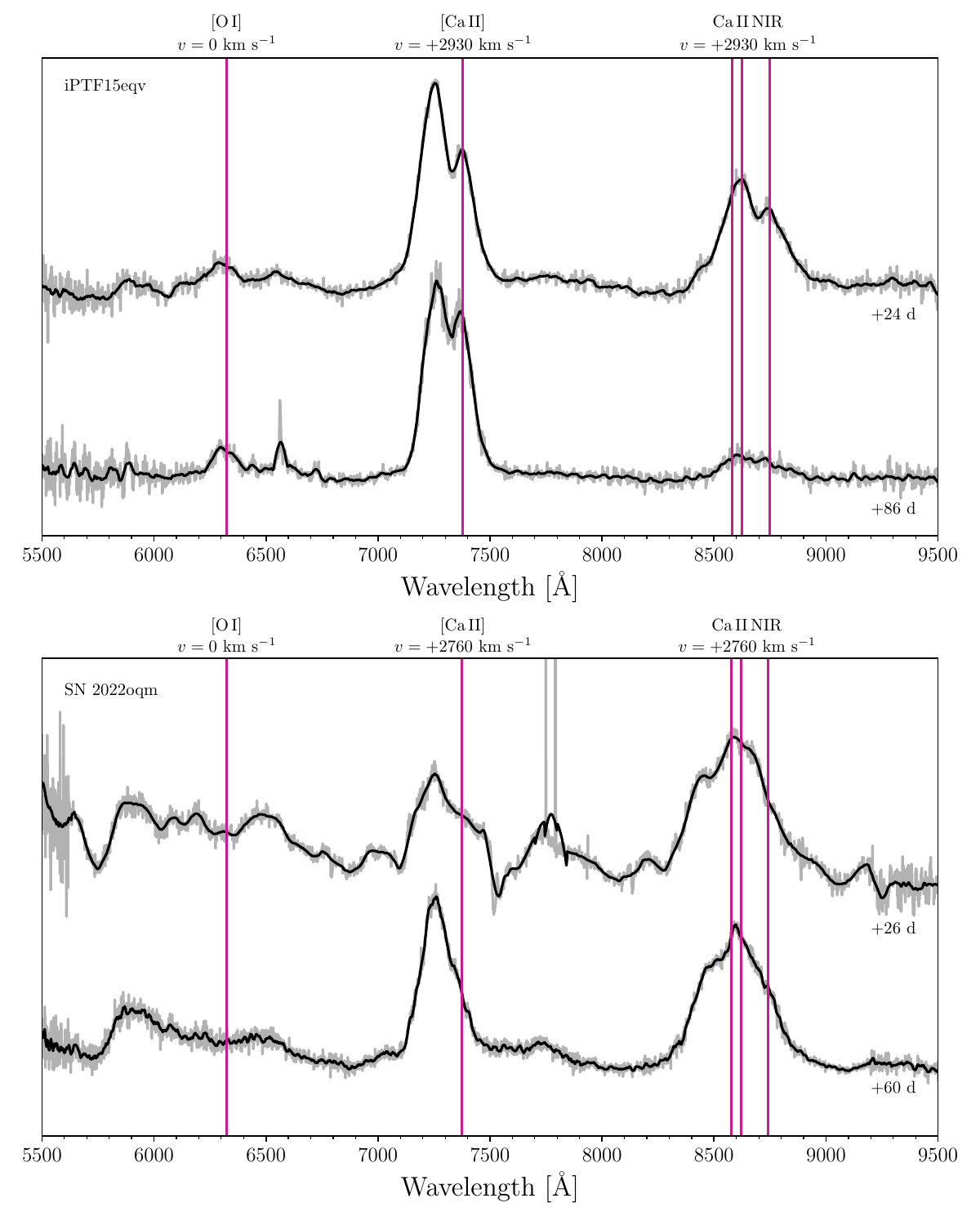}
    \caption{Two spectral epochs of objects iPTF15eqv (upper) and SN~2022oqm (lower). Pink vertical lines show the rest wavelength of the [O~I] emission feature and the [Ca~II] and Ca II NIR features at the velocity associated with the `low velocity' component. For both objects, their significantly red-shifted `low-velocity component' [Ca~II] emission aligns with the velocity of the Ca II NIR feature.}
    \label{fig:eqv_oqm_neb_vel}
\end{figure}

A handful of transition features were still observable well into the traditionally purely `nebular' regime, with the transition features of 6 objects (SN~2003H, iPTF15eqv, SN~2018gwo, SN~2018kjy, SN~2019ofm, and SN~2019pof) remaining observable until well after +100 d. This could suggest that these have a stronger contribution from a region external to the SN, allowing it to be seen until late times. This is discussed further in Section \ref{prog_forb_trans}. This persistent contribution from the `high-velocity' component deep into the nebular regime may contribute significantly to the overall integrated line flux of the [Ca~II] emission feature at these late epochs. As these are the spectra from which we determine the [Ca~II]/[O~I] ratio of each object -- a key diagnostic tool in their classification -- it is therefore plausible that these ratios are overestimates of the `true' ratio of emission fluxes produced by the core of the SN. The contribution from the `high-velocity' component at these late epochs is, however, minimal as the `low-velocity' component tends to dominate the line profile. In the latest available spectrum of SN~2018kjy, which has the most prominent `high-velocity' component persisting into the nebular regime, the `high-velocity' component contributes $\leq25\%$ flux to the overall emission profile. Additionally, as the [Ca~II]/[O~I] ratios of these objects with persistent `high-velocity' components are comfortably above the threshold value of 2, we are confident each of these objects are true members of the Ca-strong class and have not been misclassified.

Of the 26 objects with spectra after the expected transitionary threshold, only four spectral epochs do not have values of \(\Delta\)AIC and \(\Delta\)BIC that indicate a preference for a two-component fit to the [Ca~II] emission feature: SN~2019ofm at +173.4~d, SN~2010et at +84.6~d, SN~2005cz at +40.0~d, and SN~2022oqm at +60~d. As we are choosing to select the most appropriate model with respect to both of these criteria, and each of these spectra are outside our expected phase range for transitional features, we therefore disregard these four spectral epochs in the analysis of the transitional feature.

Figure~\ref{fig:ca_x2} shows the velocity and full-width at half maximum (FWHM) evolution with time of each of the component velocities in each object. The `high-velocity' and `low-velocity' labels are relative for a given SN -- the high-velocity feature is defined as the component of the emission feature with the greatest blue-shifted velocity, and the low-velocity feature is that with the least blue-shifted velocity. The velocities of both the high-velocity and low-velocity components of the [Ca~II] emission features remain constant within  3\(\sigma\).  For objects iPTF15eqv and SN~2022oqm, their red-shifted `low-velocity' components are investigated in Figure~\ref{fig:eqv_oqm_neb_vel}. The [O~I] emission in both objects is consistent with a velocity of 0 km~s\textsuperscript{$-$1}. Both the Ca~II NIR~\(\lambda\lambda\)8498,8542,8662 and [Ca~II] features, however, are instead consistent with the mean measured velocity of the `low-velocity' component of their respective objects (Figure~\ref{fig:ca_x2}), indicating the red-shift we observe is not unique to the [Ca~II] emission feature, but an offset in both nebular Ca emission features.

The FWHM of the high- and low- velocity components remain constant as each object evolves to within 3\(\sigma\), with the exception of SN~2022nwx. Determining the true FWHM at each epoch requires us to accurately identify the start and end wavelength of each component of the [Ca~II] emission line, which is difficult due to the complex nature of the photospheric phase spectral continuum. Additionally, this analysis assumes the only contributing flux to each emission component is from [Ca~II] emission; any contribution to these line profiles as a result of blended features with other species will artificially broaden the feature. Although, as discussed, we do not expect significant contamination from other features. As a result, the uncertainties of the FWHM measurements are significantly higher at earlier epochs. Additionally, the anomalous behaviour of SN~2022nwx may be attributed to a misidentification of the start and end points of its low-velocity component in the earlier epochs. Despite this source of increased uncertainty, the evolution -- or lack thereof -- of the FWHM of each component provides valuable insight into the behaviour of Ca-strong SNe. A constant FWHM through time points towards emission from a region which is fixed in velocity-space. We can therefore infer each of the emitting regions, from which our distinct emission components may arise, are fixed in velocity-space.

\section{Photometric analysis of early-flux excess}
\label{lighty_ppb}

Previous investigations into the early photometric behaviour of Ca-strong objects have shown they can display an early flux excess (or pre-peak bump) \citep{jacobson2020sn, jacobson2022circumstellar}, with distinct deviation in their early-time light curves from the rise of the main, \textsuperscript{56}Ni-powered, peak.  In our full sample of 46 Ca-strong objects, 26 objects were photometrically observed between $-$16 and $-$5~d from peak light, and as such were included in this section of the analysis. A preliminary classification of the early flux excess was determined by visually inspecting the light curves for our sample for deviations from a smooth rise between $-$16 and $-$5 d. Objects with at least two photometric observations in one band that deviated distinctly from an otherwise smooth rise were labelled as `present bump' objects, those with no clear deviation as `no bump' objects, and any objects that were not sampled at a high-enough cadence pre-peak to determine this -- or the presence of a bump was ambiguous -- as `possible bump' objects. No object in our sample had pre-peak data of a high enough cadence to definitively rule out the presence of a bump. Of these 26, 15 objects were categorised into the `present bump' category: SN~2018kjy, SN~2018lqo, SN~2018lqu, SN~2019ehk, SN~2019hty, SN~2019ouq, SN~2019pof, SN~2019txr, SN~2021acnz, SN~2021gno, SN~2021inl, SN~2021sjt, SN~2022nwx, SN~2022oqm, SN~2023er. The remaining 11 objects were classed as `possible bump' objects based on this initial visual inspection: SN~2018ckd, SN~2018dbg, SN~2018kqr, SN~2019gau, SN~2019ofm, SN~2019ouq, AT~2021zfp, SN~2023er, SN~2024jgw, SN~2024uai, AT~2024aaab.  These preliminary categories guided the initial analysis technique applied to each object; `present bump' objects were modelled using the methodology outlined in Section \ref{sce_present}, and `possible bump' objects with that described in Section \ref{sce_mod_lims}.

\subsection{Early flux excess in other SNe}
\label{other_ppbs}

Early flux excesses have been observed in various types of SNe. In Figure~\ref{fig:bumps_iib_comparison}, we display the $r$-band light curves of each of our `present-bump' objects, compared to the light curves of three other distinct SNe with known pre-peak emission -- SN~2017cbv \citep[SN Ia,][]{2017ApJ...845L..11H}, iPTF15dtg \citep[SN Ic,][]{2016A&A...592A..89T}, and SN~2011fh \citep[SN IIb,][]{2022ApJ...928..138P}. Each subplot displays one of our objects in direct comparison to these comparison objects, scaled to the peak brightness of each Ca-strong object. In this section of analysis, we are not assuming that the physical origin of the early flux excesses are the same between Ca-strong SNe and these other classes exhibiting an early flux excess, but are instead using these comparisons to constrain the differences in occurrence time and duration of the flux excesses.

Across our Ca-strong sample with these early bumps, there is clear deviation in the behaviour of their photometric data and those of the comparison objects. The first deviation is present in the decline rate and shape of the main peak of the light curves: the Ca-strong sample decay much faster after peak than any of the three comparison objects, resulting in narrower light curves. Notably, SN~2019ehk, SN~2019pof, and SN~2021sjt -- which were previously classified as SNe IIb-Ca-strong \citep{2023ApJ...959...12D} -- show significantly faster light curve evolution than the Type IIb comparison event, serving as further evidence that they are likely `true' members of the Ca-strong class, as discussed in Section~\ref{ztf_sample}. Additionally, the comparison SN Ia and SN Ic objects have much less prominent PPBs than the Ca-strong objects that exhibit this phenomenon -- the PPB of SN~2017cbv presents more-so as a broadening of the light curve at early-time as opposed to a distinct rise-and-decline behaviour as seen in many Ca-strong objects (e.g. SN~2019ehk, SN~2019ouq, SN~2021sjt, etc).  

Another clear distinction between the light curves of these comparison objects and the Ca-strong events are in the behaviour of the PPB itself. The timescale of the PPBs in Ca-strong SNe appear to occur over shorter timescales than in the other classes of SNe. While the exact duration of each PPB in the Ca-strong sample cannot be constrained, in well-sampled objects such as SN~2019ehk, we see that Ca-strong early flux excesses typically have a timescale of \(\sim\)5~d from start to finish. The comparison SNe, however, have much longer timescales of this early flux excess: the decline of SN~2011fh (as its full early flux excess behaviour was not observed) lasts $\leq$6 d, the PPB of SN~2017cbv extends over \(\sim\)10~d, and the PPB timescale of iPTF15dtg is constrained to $\mathrm{t}\geq10$~d -- each implying larger timescales than any observed PPB across our Ca-strong sample.

\subsection{Shock cooling emission model}
\label{SCE_time}

\cite{2015ApJ...808L..51P} describe a generalised prescription of SCE and provide a time-dependent light curve model to describe the first peak of double-peaked light curves, which we apply to our sample. It has been previously used to model the high-cadence double-peaked early time light curve of SN~2019ehk \citep{jacobson2020sn}. This SCE model approximates the luminosity produced when an explosion shock deposits energy into a low-mass extended envelope of mass $M_{\rm e}$, and radius $R_{\rm e}$, surrounding a core of mass $M_c$.  The SCE model is a one-zone model; it assumes a constant density in the homologously-expanding and spherically-symmetric extended material and is agnostic to its specific chemical composition, and thus is especially well-suited to the unknown progenitor composition of Ca-strong SNe. These assumptions limit the extent to which the derived envelope masses and radii can be disentangled, thus the derived values in this section of analysis can only serve as estimates on the `true' values of the envelopes surrounding each Ca-strong object.

Following the analytic treatment of \cite{2008MNRAS.383.1485V}, as discussed by \citet{jacobson2020sn}, the ejecta mass, $M_{\mathrm{ej}}$,  and kinetic energy, $E_{\mathrm{k}}$, of a SN are related by
\begin{equation}
M_{\mathrm{ej}} = \frac{10}{3} \, \frac{E_{\mathrm{k}}}{v^2},
\label{mej_degen}
\end{equation}
where \textit{v} is the photospheric velocity of the ejecta. Ca-strong SNe typically have been found to have $M_{\mathrm{ej}}$ of 0.1 $-$ 1.5 M\textsubscript{\(\odot\)}, and \textit{v} of 5000 $-$ 11000 km~s\textsuperscript{$-$1} \citep{de2020zwicky}. Using these ranges in Eqn.~\ref{mej_degen} yields an expected $E_{\mathrm{k}}$ range of 1.5\(\times\)10\textsuperscript{49} $-$ 1.1\(\times\)10\textsuperscript{51} erg.

\subsection{SNe with pre-peak bumps}
\label{sce_present}

Our implementation of the SCE model computes the bolometric luminosity of the model, converts it to band-specific fluxes, and performs MCMC fitting to our photometric data using the \texttt{emcee} ensemble sampler to infer the physical parameters  \citep[$M_{\rm e}, R_{\rm e}$;][]{2013PASP..125..306F}. We used a fixed core mass of 0.6 M\textsubscript{\(\odot\)}. As a test of our model implementation, we fit the light curves of SN~2019ehk using the same opacity ($\kappa = 0.2\ \mathrm{cm^2\ g^{-1}}$) and SN energy ($E_{\mathrm{SN}} = 1.8\times 10^{50}\,\mathrm{erg}$) as in \cite{jacobson2020sn}. We find best fit parameters of $M_e = 0.006 \pm 0.001\,M_\odot$, and $R_e = 146 \pm 22\,R_\odot$, consistent with the values of \cite{jacobson2020sn} to 1.3\(\sigma\) despite a changed value of core mass from 1 to 0.6 M\textsubscript{\(\odot\)}. 

\begin{figure}
	\includegraphics[width=\columnwidth]{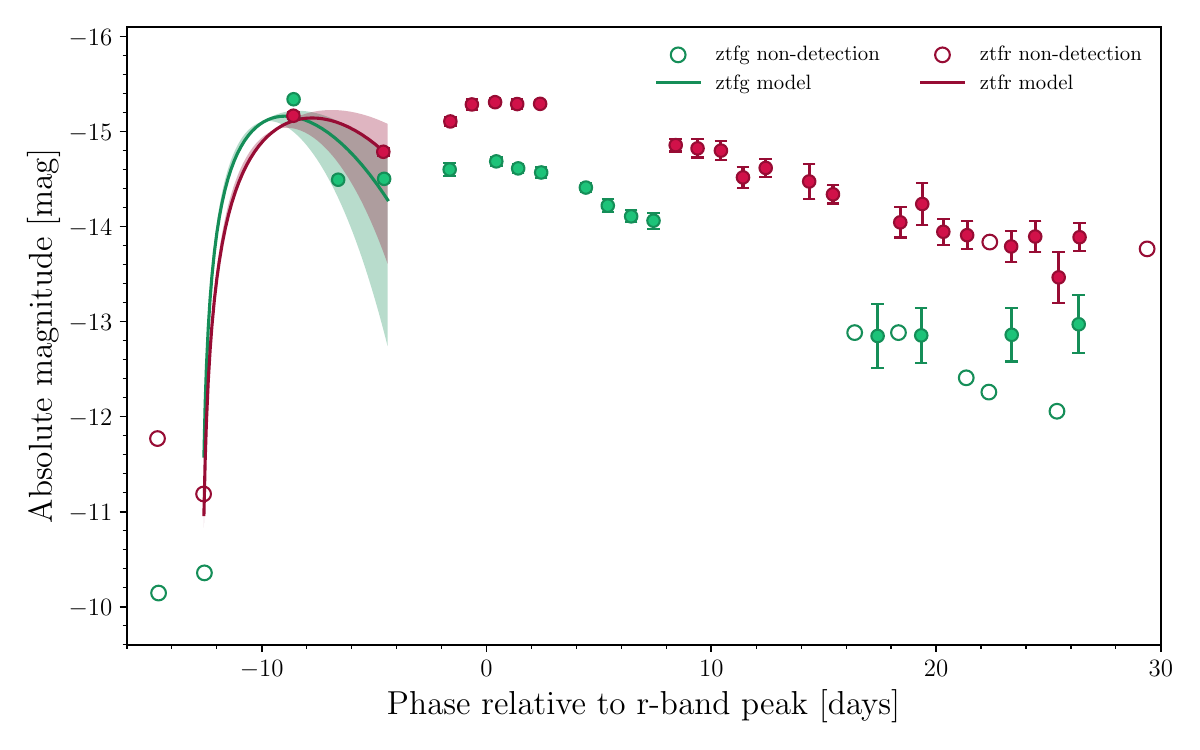}
    \caption{Early flux excess SCE fit to SN~2022nwx. Solid points show ZTF data at a significance level of $>$3\(\sigma\), and open points show non-detections for ZTF $g-$ (green) and $r-$ (red). The $g-$ and $r-$band models are fitted simultaneously, and are shown in red and green solid lines, respectively, with their associated 1\(\sigma\) uncertainties shaded.}
    \label{fig:nwx_fit}
\end{figure}

To investigate early flux excesses in our Ca-strong SN sample, we applied this analytic SCE model to the observed photometry of each event. All photometric measurements from ZTF and literature sources were binned nightly prior to fitting. For photometry of literature objects which were only available in magnitudes, we converted these to comparable flux units. We defined a `non-detection' as any measured flux with a significance of less than 3\(\sigma\). To ensure that only the early flux excess was modelled with this SCE approach, we restricted the fit to data between the latest non-detection prior to first light and the onset of the main light curve rise for each SN (typically $-$16 to $-$5~d pre-peak) -- this value was chosen manually per SN.

Using a grid search combined with MCMC sampling, we fitted the model simultaneously to the multi-band light curves of each event at 15 fixed explosion energies between 1.5\(\times\)10\textsuperscript{49} $-$ 1.1\(\times\)10\textsuperscript{51}~erg as derived above, determining the best-fit energy from the associated model with a reduced chi-squared value closest to 1. Given the well-constrained nature of the SN~2019ehk early light curve fits, we used its fitted SCE parameters as our prior centroids in the fitting for the rest of the `present bump' sample. We used logarithmic priors with ranges of $M_{\mathrm{e}} = 0.01-0.51\,M_\odot$ and $R_{\mathrm{e}} = 10-5000\,R_\odot$. Upper limits of these priors were chosen based on the largest model values of \cite{2015ApJ...808L..51P}, while the lower limits were chosen as 0.02 times the smallest model values of \cite{2015ApJ...808L..51P} to account for the comparatively lower luminosities of Ca-strong SNe.

The posteriors from our MCMC sampling provided marginalised constraints on $M_{\rm e}$ and $R_{\rm e}$. We took the median values and 16\textsuperscript{th} and 84\textsuperscript{th} percentile intervals as the estimated parameter value and its associated 1-\(\sigma\) uncertainties, respectively. We generated synthetic light curves using the best–fitting parameters, and compared them directly to the binned multi–band photometry. Figure~\ref{fig:nwx_fit} shows an example of the best fitting light curve for SN~2022nwx in our `present bump' sample.

\subsection{SNe with possible pre-peak bumps}
\label{sce_mod_lims}

\begin{figure}
	\includegraphics[width=\columnwidth]{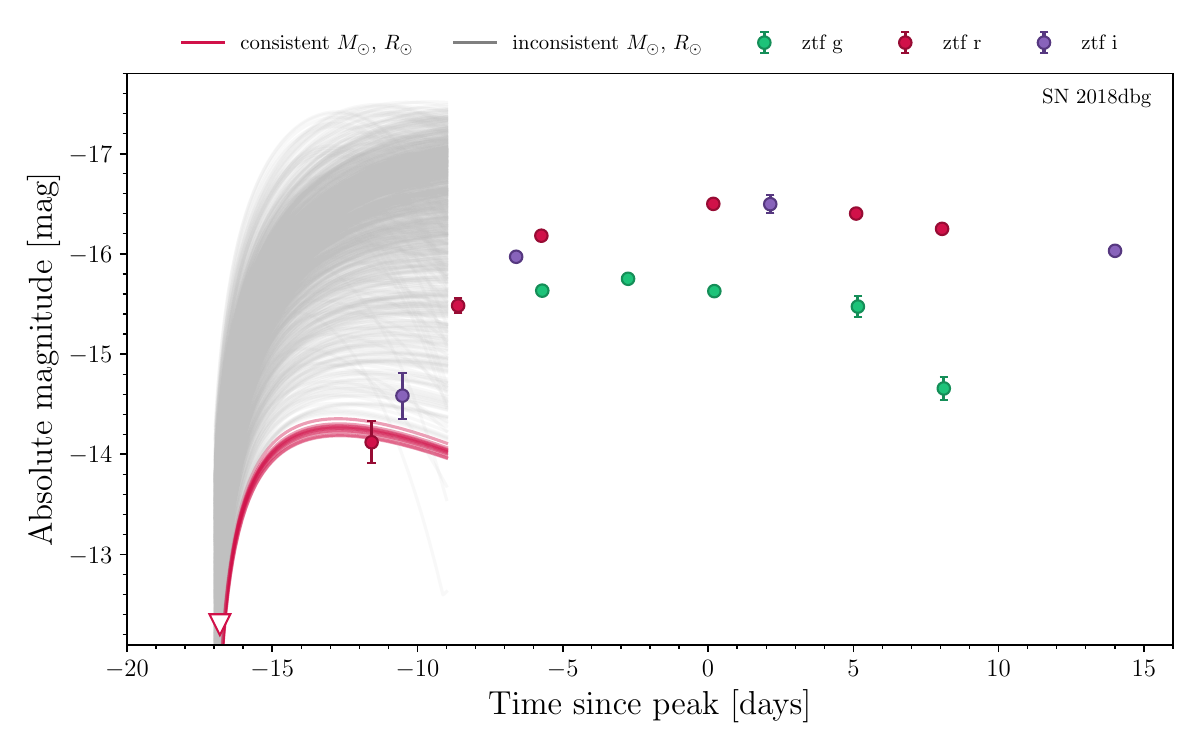}
    \caption{SCE model fitting procedure to determine the upper limits of M\textsubscript{e} and R\textsubscript{e} for a `possible bump' object -- SN~2018dbg. Green, red, and purple points show the ZTF \textit{g-}, \textit{r-}, and \textit{i-} band photometric observations. Grey lines are SCE models of M\textsubscript{e} and R\textsubscript{e} parameter combinations which are inconsistent with the observed r-band photometric data. Red lines are SCE models which are consistent with the $r-$band observations.}
    \label{fig:ppb_ambig_fits}
\end{figure}

The SCE models for each of the 15 events in our sample with confirmed bumps were used to evaluate whether an early flux excess could be ruled out in our `possible-bump' sample of 11 objects. For each object, the range of allowed pre-peak fluxes from our SCE model outputs was compared to the observed light curve. We adopted a criterion that an early bump would be considered undetected if the predicted flux does not exceed $3\sigma$ above the local photometric uncertainty. Objects with upper limits below the predicted flux of all models were considered as not showing a bump, whereas objects with non-constraining limits were flagged as potential SCE candidates. No objects in our sample were originally visually classified as `no bump' objects, nor did they satisfy the criterion required to be categorised as `no bump' objects here, and as such all remaining Ca-strong SNe in our sample were deemed to be `possible bump' objects.

For the objects flagged as `possible bump' candidates, we determined the upper limits of M\textsubscript{e} and R\textsubscript{e} that were consistent with each object's observed photometry. Figure~\ref{fig:ppb_ambig_fits} shows the SCE model fitting procedure we used to determine the upper limits of M\textsubscript{e} and R\textsubscript{e} for a `possible bump' object -- SN~2018dbg.  Taking the SCE model with the highest flux consistent with the photometry -- the red line with the brightest absolute magnitude in Figure~\ref{fig:ppb_ambig_fits} -- gives the model with the highest M\textsubscript{e} and R\textsubscript{e} consistent with observations. These values of M\textsubscript{e} and R\textsubscript{e} therefore serve as the upper limit estimates for this object. If any objects had data of a high enough cadence to constrain an SCE model fit further than just an upper limit -- i.e., more than two photometric data points describing the rise and decline of the PPB in any given band -- we re-classified these into the `present bump' sample. This was the case for objects SN~2019hty and SN~2019txr, which were added to the `present bump' sample, and the full range of allowed pre-peak fluxes from our SCE model outputs (as described above) was re-determined. As these objects had values of M\textsubscript{e} and R\textsubscript{e} consistent with the rest of the `present bump' sample, their inclusion into this sub-sample did not alter our previously-determined range of expected values of M\textsubscript{e} and R\textsubscript{e} in Ca-strong objects.

If the time between the last non-detection and first 3\(\sigma\) detection was less than 10 days, we classed the model as consistent with the photometry if the model and observed data were in agreement within 1\(\sigma\). For those with no photometry within this range, we could neither rule-out the presence of a PPB, nor constrain any physical parameters of the explosion environment. We randomly sampled a grid of physically plausible M\textsubscript{e} and R\textsubscript{e} values, derived from the fits of the sample with clear early flux excesses, and derived the associated \cite{2015ApJ...808L..51P} models associated with each combination of values for a fixed E\textsubscript{SN}. E\textsubscript{SN} was fixed at the lowest value preferred across the objects with a confirmed bump -- 3.31\(\pm\)0.03\(\times\)10\textsuperscript{50} erg s\textsuperscript{$-$1}. Using this value of lowest plausible SN energy ensured the M\textsubscript{e} and R\textsubscript{e} values derived from this analysis were true upper limits on the ejecta required to reproduce the observed behaviour.

\subsection{SCE fitting results}

\begin{figure}
	\includegraphics[width=\columnwidth]{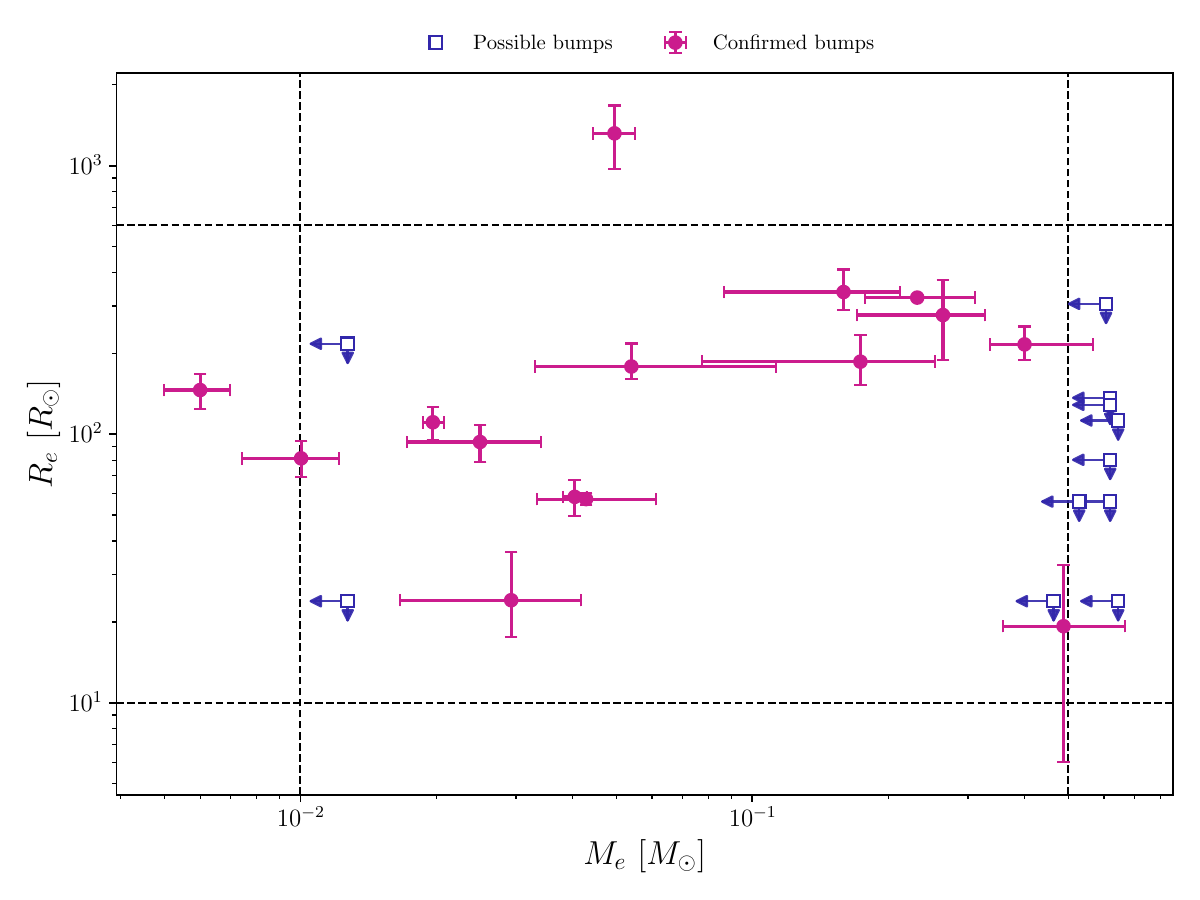}
    \caption{Parameter space of envelope mass and radius for all `present bump' and `possible bump' objects across our sample. Pink solid circles show the derived M\textsubscript{e} and R\textsubscript{e} for `present bump' objects. Open blue squares represent the upper limits of M\textsubscript{e} and R\textsubscript{e} associated with each of the `possible bump' objects. The dashed lines denote the range of values expected for typical SNe IIb. }
    \label{fig:pspac}
\end{figure}

While the SCE modelling can confidently reproduce this early photometric phenomenon observed in many Ca-strong objects in our full sample, there is still clear diversity between each of these objects. Figure~\ref{fig:pspac} shows the M\textsubscript{e} against R\textsubscript{e} parameter space of these model fits of our sample of Ca-strong events. The dashed lines denote the range of values expected for typical SNe IIb -- the most prominent subclass of object with early flux excesses \citep{nakar2014supernovae, arcavi2017constraints}. Each of the `confirmed bump' parameters lie within 3\(\sigma\) of the range of values for SNe IIb. The PPBs in our sample (see Figure~\ref{fig:bumps_iib_comparison}) show a broad range of behaviours, with peak brightnesses of between $-14.32\pm0.04$ (SN~2021gno) and $-$17.33$\pm$0.02 mag (SN~2022oqm), durations of \(\sim\)5~d (SN~2019ehk) to \(\sim\)9~d (SN~2022oqm).

A key caveat to the methodology we have employed is that it assumes the extended material is symmetric and spherical. Any asymmetries in the progenitor system would vary both the mass and radius along the line of sight and could significantly impact the luminosity of a `bump' due to SCE of this external material.

Among the sample of objects with higher-cadence photometric observations across this typical phase range of Ca-strong early flux excesses (i.e. $>$2 distinct epochs between last non-detection and first 3\(\sigma\) detection), each object fell into the `present bump' category. Without high-cadence data within this critical range of $-$16 and $-$5~d in which we would expect a PPB to occur, it is impossible to fully constrain the behaviour of any early flux excess in the `possible bump' sub-category, or to rule out the presence of a bump entirely. As Ca-strong objects are intrinsically faint, they often are not detected until past this critical range of phases. Despite these observational challenges, over half of the objects from our total Ca-strong sample are observationally consistent with an early flux excess. 33\% of objects in our sample fall into the `present bump' sub-class, with a further 24\% defined as `possible bump' objects; the remaining objects are unfortunately excluded from this area of analysis due to insufficient data (see Section \ref{lighty_ppb}). As such a significant proportion of these objects show this behaviour, and we cannot definitively rule out the presence of a bump in any Ca-strong object in our sample, we may speculate that these PPBs may be a phenomenon common across this class of transient.

\section{Discussion}
\label{hi ca and o}

\subsection{Widespread detection of early forbidden features}
\label{early_ca_o}

One of our primary objectives was to identify and characterise, in a large sample of Ca-strong SNe, any early [Ca~II] and [O~I] emission, such as that observed in the Ca-strong object SN~2023xwi \citepalias{2025MNRAS.537.1015T}. From our full sample of 46 Ca-strong SNe, 35 objects had the required spectra within 7~d of peak, and of these 35, 29 had measurable forbidden [Ca~II] emission in this phase range. If we extend this `peak' cut-off to 10~d, this increases to 32 of the 35 total objects (91\%) having early forbidden [Ca~II] emission. The physical environment responsible for forbidden emission features, traditionally only seen at late times due to the low-density requirements for its associated atomic transition, must therefore be common in Ca-strong SNe. 

Forbidden emission typically emerges once the ejecta have expanded, cooled, and become optically thin, and the photosphere itself has receded. This observed early time forbidden emission is therefore difficult to reconcile with an intrinsic origin, as the ejecta are expected to be too dense at these phases for strong forbidden emission to occur. 

The NLTE CO+He WD model of Ca-strong events presented by \cite{2025A&A...702A..29C} produces distinct [Ca~II] emission by +20~d -- 10~d earlier than the typical expected threshold phase for the emergence of nebular emission lines in Ca-strong objects. This is, however, still at a considerably later phase than we observe across the sample of these events, with 91\% of objects displaying distinct [Ca~II] emission within +10~d of peak light. While these results demonstrate that forbidden emission can occur earlier than traditionally expected, the observed early-time spectroscopic behaviour of Ca-strong SNe is still yet to be reproduced by models for Ca-strong SNe.

Peculiar ejecta structures, such as steep density gradients, may permit the emergence of forbidden emission lines at earlier epochs than typically expected \citep[e.g.][]{{2025A&A...699A.169B}}. While we therefore cannot rule out an intrinsic ejecta origin for the early [Ca~II] emission observed in Ca-strong SNe, such behaviour has yet to be reproduced around peak light across the full population of these objects. We instead favour an interpretation in which the observed early-time [Ca~II] emission originates from an external, low-density region of Ca that predates the SN explosion, naturally providing the conditions required for early forbidden emission.

The same is true for the observed [O~I] emission, indicating that this external region of matter must, at least, include a combination of low density Ca and O. When originating from the same emitting region, [Ca~II] emission typically dominates any [O~I] emission as cooling via [Ca~II] emission is far more efficient than through [O~I] -- the Einstein A coefficient of [Ca~II] is two orders of magnitude greater than that of [O~I] \citep{jacobson2022circumstellar}. To observe both forbidden emission lines concurrently, we therefore require either substantially higher abundances of O than Ca, or the emitting regions to not be co-located. The range of non-zero velocities around peak light of the [Ca~II] and [O~I] emission in the sample of Ca-strong SNe places further constraints on the external, pre-existing, environment of these objects and is likely the result of different viewing angles, thus implying that the potential progenitor systems are inherently aspherical.

\subsubsection{Forbidden velocity evolution and transition features} 
\label{vel_ev_dist}

As discussed in Section~\ref{early_ca_o}, the early forbidden [Ca~II] and [O~I] emission lines likely originate from an external low-density region but as the ejecta expand and cool and the photosphere recedes, we expect the [Ca~II] and [O~I] features to instead be dominated by the core SN emission at late times. This would present itself as an evolution in velocity space from a non-zero velocity around peak to a zero velocity at late-times. This evolution of velocity is seen in our SN sample (see Figure~\ref{fig:all_vels_dist}), indicating that for Ca-strong SNe, the source of forbidden emission may transition from a low density external region in the photospheric phase, to a more core-dominated emission regime in the nebular phase.

Figure~\ref{fig:ca_x2} shows the velocity and FWHM evolution of each component velocity for each of the SNe with observed transition signatures in the [Ca~II] emission profile. Both the high- and the low-velocity components of the [Ca~II] emission line for each SN remain constant with time. We suggest that these high-velocity and low-velocity components are the physical counterparts to the external-structure-dominated emission region (high-velocity component) and the core-dominated emission region (low-velocity component). This indicates that the line profile evolution we see in each of these objects, such as that seen of SN~2023xwi in Figure~\ref{fig:example_model_fit}, is not a result of the deceleration of any external, pre-existing material as the SN enters the nebular regime. Instead, the profiles we observe may be indicative of the relative increase of core emission, which ultimately dominates over the external emission in the nebular regime, resulting in our final observed velocities of these forbidden lines that are broadly consistent with zero velocity. As shown in Figure~\ref{fig:eqv_oqm_neb_vel}, for the objects with an offset `low-velocity' emission component, this apparent red-shifted velocity is consistent with the velocity of the NIR triplet emission. Therefore, while their `low velocity' component is not necessarily consistent with 0 km~s\textsuperscript{$-$1}, it is still consistent with the nebular Ca emission associated with the SN.

Currently, rapid classification of candidate Ca-strong events for follow-up relies on criteria such as a low absolute magnitude and a H-free spectrum. Incorporating searches for early-time [Ca~II],  and in most cases [O~I], which is typically detected within 7~d from first [Ca~II] detection, could significantly improve early classification. This would allow more targeted follow-up, ensuring fewer observational resources are spent on candidates that ultimately prove not to be Ca-strong.

\subsection{Early flux excesses in all Ca-strong SNe?}
\label{ppb_disc}

In our full sample of 46 Ca-strong objects, 15 had distinct early flux excesses in their light curves. When we include the further 11 objects with at least one photometric point that deviated from a smooth rising light curve, all 26 Ca-strong objects with pre-peak observations are consistent with an early flux excess. Additionally, there is no SN in our sample for which we can definitively rule out the presence of an early flux excess. The `present bump' and `possible bump' objects span the full range of observed peak brightnesses and redshifts of the full Ca-strong sample, suggesting the presence of a PPB is not magnitude limited, and detection is instead dependent on early-time sampling of the light curves. The high fraction of Ca-strong objects in which we see a PPB, despite the observational challenges, indicates that they may be a common trait across all Ca-strong SNe. 

We find timescales for the PPB of 5 -- 9 d and peak brightnesses of $\sim$ $-14.3 - -17.3$ mag. In Figure~\ref{fig:pspac} we presented the parameter space of envelope mass and envelope radius for all 26 objects in our sample of Ca-strong SNe for which we could constrain the SCE model, or estimate upper limits on the parameters, that would produce an SCE model consistent with our observations. We found envelope masses of 0.006 -- 0.54 $M_{\odot}$ and envelope radii of 24 -- 1320 $R_{\odot}$ for those that could be modelled. 

\begin{figure}
	\includegraphics[width=\columnwidth]{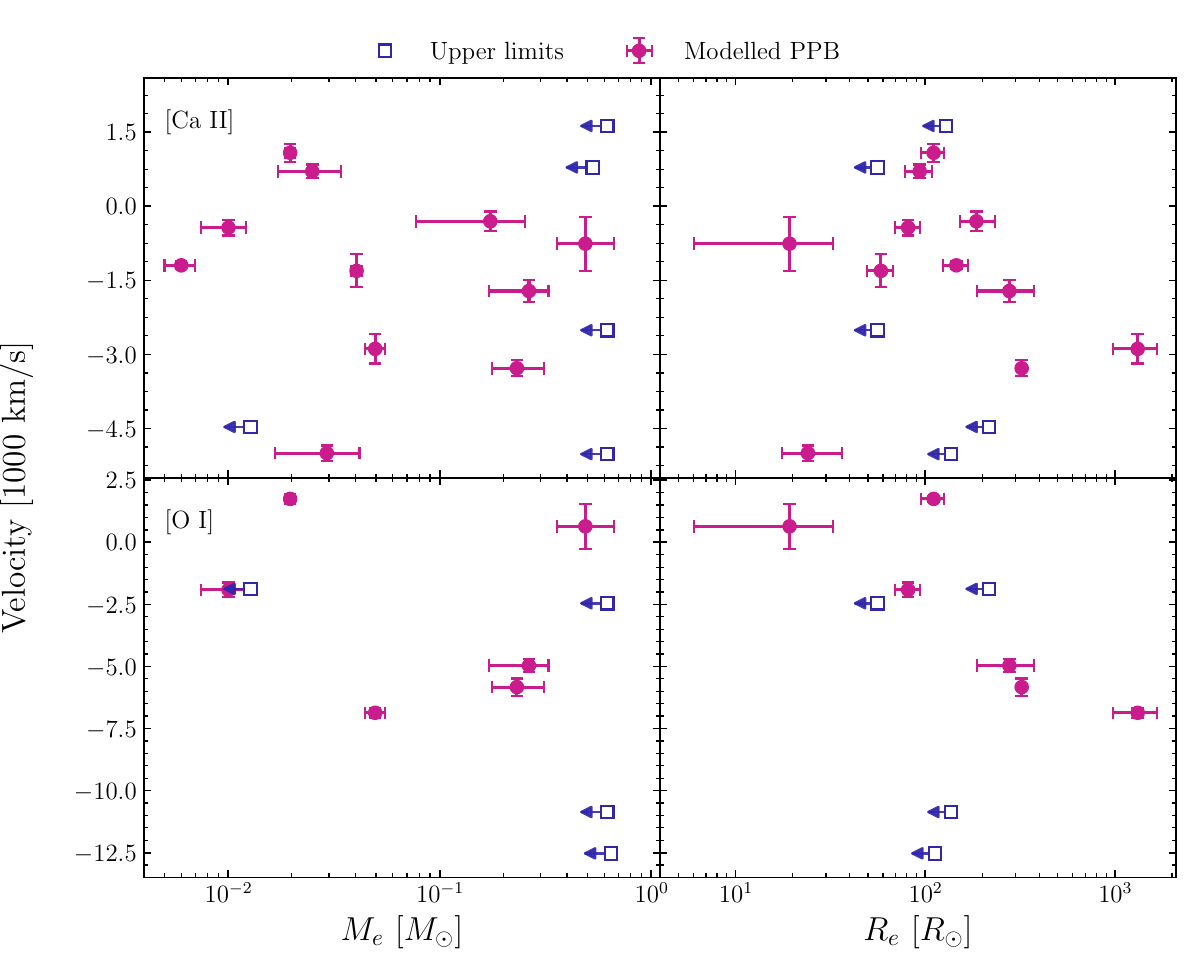}
    \caption{Velocity at peak of forbidden [Ca~II] (upper) and [O~I] (lower) and envelope mass (left) and radius (right) required to reproduce observed photometric early flux excess behaviour. Pink circles show values associated with the `present bump' sample, and blue open squares show the upper limits associated with `possible bump' objects.}
    \label{fig:mass_rad_vel}
\end{figure}

To explore potential physical connections, we investigated correlations between early flux excesses and spectroscopic features, including the velocities of early forbidden lines, as well as other photometric properties. Comparisons between the `present bump' and `possible bump' populations were made to assess whether they represent distinct subgroups within the Ca-strong class. Figure~\ref{fig:mass_rad_vel} shows the [Ca~II] (upper) and [O~I]~\(\lambda\lambda\)6300,6364 (lower) velocities against the derived envelope mass and radius values. There is no obvious correlation between the early forbidden emission line velocity, and either the envelope mass or radius. Due to the intrinsic degeneracy between M\textsubscript{e} and R\textsubscript{e} in\, and thus the limitations of, the \cite{2015ApJ...808L..51P} model, it is not unexpected that no such correlations were found. Future modelling of Ca-strong SNe and their progenitor environments which are not dependent on the same assumptions of spherical symmetry and a constant density profile may reveal a correlation between these parameters.

\subsection{Ca-strong sub-classifications}
\label{sub_classificaysh}

\cite{de2020zwicky} introduced further sub-classification of Ca-strong objects: Ca-strong-Ia and Ca-strong-Ib/Ic. This was based on the line depth of the Si~II~\(\lambda\)6150 feature, \(a\)\textsubscript{Si~II~\(\lambda\)6150} at peak light, as defined by \cite{sun2017quantitative}. They showed that `normal' SNe Ia and SNe Ib/Ic are split \citep[see figure~9 of][]{sun2017quantitative}, with $a_{\mathrm{Si\,II}\,\lambda 6150} > 0.35$ for SNe Ia and $a_{\mathrm{Si\,II}\,\lambda 6150} \leq 0.35$ for SNe Ib/c. Using the same method, we measured this for all Ca-strong objects in our sample with spectra around peak light with a resolution of $\leq$20 $\AA$ and SNR $\leq$3 to ensure correct identification and accurate modelling of the feature. From our full sample of 46 confirmed Ca-strong objects, 18 objects did not meet these criteria. The resulting values for the remaining 28 objects are shown in Table \ref{tab:ca_rich_line_depths}. In addition to the sub-classification of \cite{de2020zwicky}, we introduce a Ca-strong-ambiguous class -- objects for which their measured $a_{\mathrm{Si\,II}\,\lambda 6150}$ and associated uncertainties are consistent with the Ia/Ibc boundary of $a_{\mathrm{Si\,II}\,\lambda 6150} = 0.35$ within 3\(\sigma\).

Of the 29 SNe with spectra around peak light (within +7~d from $r$-band maximum), only one object  -- SN~2016hnk -- is definitively Ca-strong-Ia. Five further objects fall in the Ca-strong-ambiguous class (iPTF09dav, SN~2018lqo, SN~2018lqu, SN~2019ofm, and SN~2021inl), and the remaining 23 objects fall into the Ca-strong-Ib/Ic classification. As SN~2016hnk does not exhibit properties that significantly differ from those of other Ca-strong SNe (i.e., values of M\textsubscript{peak}, $\Delta$m\textsubscript{7}(r), v\textsubscript{[Ca~II], peak}, v\textsubscript{[O~I], peak} are all comparable to that of the rest of the sample), there is no strong justification for subdividing these events into Ca-strong-Ia and Ca-strong-Ib/Ic sub-classes. Additionally, such sub-categorisation can imply that `Ca-strong-Ia' and `Ca-strong-Ib/Ic' have different physical origins. The continuous range of spectroscopic and photometric behaviours we observe across our full sample of Ca-strong SNe, however, along with the remote origins of many being incompatible with core-collapse origins \citep{lyman2014progenitors, dong2022physical}, make such a distinction misleading. We therefore disregard this avenue of sub-categorisation of Ca-strong SNe, and instead assume these objects are not comprised of distinct heterogenous subclasses.

\section{Implications for the progenitors of Ca-strong SNe}
\label{puppis_implicaysh}

As discussed in previous sections, the early presence of [Ca~II] and [O~I] emission lines, along with the velocity evolution of these features, point towards Ca-strong progenitors occurring in environments with low-density, external regions comprised of (at least) Ca and O. Early forbidden emission lines have also been observed in the super-luminous SN, SN~2019szu \citep{2024MNRAS.52711970A}; this object showed pre-maximum [O~II] emission which was attributed to an O-rich CSM produced by mass-loss episodes shortly prior to explosion. The prominent forbidden emission we observe across the sample of Ca-strong SNe is therefore indicative of Ca-strong SNe occurring in similarly-polluted environments. 

Modelling of Ca-strong SNe and their potential progenitors have shown their mid-to-late time behaviour (i.e. after 14d) can be reproduced by interacting He+C/O WD systems \citep{zenati2023origins, 2024A&A...683A..44M, 2025A&A...702A..29C}. The presence of forbidden emission around peak light, however, is still yet to be reproduced without prior pollution of the surrounding environment. \citetalias{2025MNRAS.537.1015T} proposed a progenitor environment polluted by a recurrent He-nova AM CVn system, based on observations of the Ca-strong object SN~2023xwi. Our analysis shows that the behaviours observed in SN~2023xwi were not unique to that specific object but are instead common across the Ca-strong class. Therefore, we investigate if this proposed origin of SN~2023xwi can be extended to describe the origins of all Ca-strong SNe.

In this proposed progenitor system, we hypothesise that Ca-strong SNe arise from recurrent He-novae AM CVn-type systems, resembling V445 Puppis \citep{2003ApJ...598L.107K, woudt2009expanding}. In such systems, a CO WD accretes He from a H-deficient companion and undergoes recurrent He-novae \citep{roelofs2007hubble}. He is swept up from the He-shell of the WD(s) and forms polar outflows along the axis of rotation of the binary system. O from the surface of the CO WD(s) also gets swept up into these lobes, creating low density O regions within the broader lobe structure. In more violent novae, small abundances of elements up to Ca can be synthesised -- these, too, get swept up into the lobes and form low-density Ca regions within this structure.  These aspherical and asymmetrically distributed low-density elements, ejected prior to the SN explosion, can explain the presence and kinematic evolution of [Ca~II] and [O~I] features in both the photospheric and nebular phases once the SN finally explodes. For a detailed schematic and explanation of this system, please see Section~6.1 of \citetalias{2025MNRAS.537.1015T}.

\subsection{Peak forbidden line emission from polar outflows from a recurrent He-nova AM CVn system}
\label{peak_line_puppis}

When the SN explodes -- before the ejecta overtake the regions of O and Ca within the lobes, and while the photosphere is still obscuring the core of the SN -- we would observe forbidden [Ca~II] and [O~I] emission from this external lobe structure. The velocities we observe of these forbidden emission line features at early times give insight into the velocities of the low-density Ca and O in the lobes. Figure~6 of \citetalias{2025MNRAS.537.1015T} shows a 2D schematic of our proposed progenitor system -- adapted from the V445 Puppis system \citep{woudt2009expanding} -- and highlights the viewing-angle dependence on the observed velocity. In such a system, the minimum velocities observed around peak correspond to observations orthogonal to the rotation axis of the WD system, and the maximum peak velocities occur when such a system is observed down the axis of the lobes. The expected observed velocity distribution in such a system should follow a broadly Gaussian distribution, with the peak falling at a non-zero velocity. Viewing the system directly down the lobe, resulting in the maximum observed velocity, will be the least likely viewing angle.

The maximum velocity we observe at peak, therefore, could correspond to the intrinsic velocity of the outflow lobes, assuming the progenitor systems of these objects are all broadly uniform. The velocity distribution of [O~I] emission -- the forbidden emission line with the highest observed velocity per object throughout our sample -- around peak light has a mean value of $-$6040 \(\pm\) 1400 km~s\textsuperscript{$-$1}. The greatest observed velocity associated with a forbidden emission line in the sample of Ca-strong objects is $-$12500$\pm$1000 km~s\textsuperscript{$-$1} \citep[SN~2022oqm, $-$8 d;][]{2024ApJ...962..109I}. While this is greater than the observed polar outflow of V445 Puppis, the only such system for which this has been measured, it is plausible that other AM CVn-type systems with greater polar outflow velocities.
In addition, due to the range of peak brightnesses reached by Ca-strong SNe, it is likely that a range of systems of varying sizes and energetics are required to explain Ca-strong SNe. The largest, and most energetic, of these systems could therefore align with the velocities we observe in e.g. SN~2022oqm.

The distribution of velocities at peak also gives us insight into the geometry of our progenitor systems. Assuming all Ca-strong SNe originate from broadly uniform progenitor systems, the observed velocity is the projected lobe velocity dependant upon the observed inclination of the system. The width of the lobes will determine the distribution of observed velocities, wherein narrow lobes would show very few objects with high velocity [Ca~II] and [O~I] features, and broader lobes would result in the majority of the [Ca~II] and [O~I] features having varied, but distinctly non-zero, velocities. The latter of these follows the observed early time distribution across our sample of Ca-strong SNe, indicating that the outflow lobes in our proposed progenitor system are likely broad.

Additionally, as cooling via [O~I] emission is relatively inefficient compared to cooling via [Ca~II] emission \citep{jacobson2022circumstellar}, to observe the distinct concurrent [O~I] and [Ca~II] emission across the sample of Ca-strong SNe, we likely require the [Ca~II] and [O~I] emission to originate from two distinct regions (see section 5.2 of \citetalias{2025MNRAS.537.1015T}). Throughout the spectral evolution of objects in our sample, only five spectra show [O~I] emission associated with a `slower' intrinsic velocity than [Ca~II] at the same epoch with a significance level of $>3\sigma$ -- in all other cases, the [O~I] emission is blueshifted to a greater velocity than the [Ca~II]. This supports the stratification of the external environment surrounding the SN, and can be explained by the pre-cursor emission of an AMCVn progenitor.

\subsection{Estimation of the `extended lobe' properties}
\label{prog_forb_trans}

\citetalias{2025MNRAS.537.1015T} hypothesised the presence of transition features in spectra with diminishing photospheric features and increasing nebular emission, assumed to start at a phase of \(\sim\)14 d. This evolution is what we observe in 22 of the 26 objects with suitable spectra in this key transitional phase.  Following our proposed progenitor system, this double-component feature is due to emission from both the lobes and the core of the SN. The presence of two distinct emitting regions at this phase provides us with two key insights into this system. Firstly, at this phase, the photosphere has not yet receded entirely -- if this were the case, the [Ca~II] feature would be entirely dominated by the zero-velocity intrinsic to the SN core. Secondly, the presence of the high-velocity component in the line profile indicates the SN ejecta have not yet interacted with the [Ca~II] emitting region of the lobes. As the emitting region in the lobes is very low density -- a requirement to produce the forbidden lines we observe -- any interaction between the emitting region of the lobes and the ejecta of the SN itself would disrupt the lobe structure in such a way that its emission would likely no longer be observable.

Therefore, we can infer the minimum size of a lobe in such a system required to reproduce the distinct lobe emission feature we observe in our Ca-strong SN sample. Using the photospheric velocity, latest epoch of observed transition feature, and rise time of the object, we can constrain the minimum physical extension of the emitting lobe. To estimate this, we therefore investigate the object for which we have observed both a photospheric spectrum, and the latest possible spectrum with a transitional feature. In our sample of Ca-strong SNe, the object for which these criteria are satisfied is PTF15eqv. This object has a photospheric velocity of 7500 \(\pm\) 120 km~s\textsuperscript{$-$1} (measured from its He~I~\(\lambda\)5876 line), and we observe its [Ca~II] transitional feature up to +152~d. Combining this with its rise time of 16~d, if we assume homologous expansion, we obtain a lower limit on the lobe size of 1.08 \(\pm\) 0.02 \(\times\) 10\textsuperscript{16} cm. The V445 Puppis system has a physical lobe extension of 1.35 \(\pm\) 0.09 \(\times\) 10\textsuperscript{17} cm \citep{woudt2009expanding}. The lower limit on the lobe size we would require for similar systems to reproduce the observed behaviour of Ca-strong SNe is therefore physically plausible.

\subsection{Early flux excesses from lobe interaction}

As discussed in Section~\ref{SCE_time}, many of the objects in our sample display an observable early flux excess before peak light, consistent with SCE from extended material surrounding the SN progenitor. Under the progenitor channel discussed here, this early flux excess emission is a result of the interaction between the SN shockwave and the polluted environment of the extended lobes of a V445 Puppis-type system. In such a system, the shockwave propagates into the lobes, and the typical SCE scenario progresses from the lobes instead of a traditional spherically symmetric envelope.

As the lobes are not spherically symmetric, the PPBs we observe would therefore also be dependent upon the observed angle of each SN event. Observing the system along the pole of the lobes would, theoretically, mean a greater `radius' (and, assuming a uniform mass distribution both radially and across our sample of Ca-strong objects throughout each lobe, a greater mass) of material shocked along the line of sight; this would, in turn, result in a brighter SCE early flux excess before peak light for any given SN. Conversely, observing the system perpendicular to the lobes would result in minimal shocked material along the line of sight, and therefore, a perhaps imperceptible PPB.

As discussed in Section~\ref{hi ca and o} the distribution of velocities of forbidden emission around peak light for any given SN may act as a proxy for the viewing angle of the progenitor system. A high forbidden line velocity at peak indicates we are observing the SN at an angle close to the pole of the lobes, and a low velocity indicates an observation perpendicular to the pole axis. It is therefore reasonable to assume that there would be a correlation between the forbidden emission line velocity at peak, and the SCE radius or mass (again assuming these two quantities are degenerate). We investigated this in Section~\ref{sce_mod_lims} and found no such correlation. However, the lack of distinct correlation between early forbidden line emission velocity and physical parameters of extended material surrounding SN progenitors is plausibly driven by the assumption of spherical symmetry in our SCE modelling; the true progenitor system is likely far more complex and asymmetrical (as discussed in Section \ref{peak_line_puppis}), thus these assumptions are over-simplifications. This, combined with the inherent degeneracy between the derived envelope parameters in the \cite{2015ApJ...808L..51P} SCE model, results in our derived values of envelope mass and radius serving as estimates on the `true' values of the environments of each object. It is therefore not a concern for the legitimacy of a V445 Puppis-type progenitor system that we do not observe a clear correlation between the forbidden line velocity at peak light and the radius and mass of the extended environment. We encourage full modelling of this proposed progenitor scenario to fully ascertain whether these predicted correlations are observed.

We argue that the combination of observed photometric and spectroscopic behaviours of Ca-strong SNe across our full sample of objects is in line with the predicted observable signatures of a recurrent He-nova AM CVn system suggested in \citetalias{2025MNRAS.537.1015T}. We can therefore postulate that such systems are responsible for all Ca-strong SNe, uniting the origins of this class of exotic transient, and providing further diagnostic tools in the early classification of future Ca-strong objects.

\section{Summary}
In this paper we have presented optical photometric and spectroscopic observations of the full sample of Ca-strong SNe. Combining literature objects with previously unclassified follow-up targets from the ZTF 2018-2024 sample (and one additional PanSTARRS object) brought our total sample up to 46 objects detected prior to 31 December 2024. Across the sample, we determined the peak brightness and decline rate from all available photometric data. An early-flux excess was distinctly present in 15 of the objects in our sample, and a further 11 objects were consistent with PPBs due to SCE. Across our full sample of 46 confirmed Ca-strong objects, we could not rule out the presence of an early flux excess in any. We investigated correlations between these objects and other observable parameters, but no significant dependencies were apparent.

The core of our investigation was focussed on the spectroscopic evolution of our sample, specifically, the evolution of forbidden [Ca~II] and [O~I] emission lines. Historically, these features have been associated with the nebular phase of these objects, their key diagnostic being a high [Ca~II]/[O~I] ratio in the nebular-phase. To observe these forbidden transitions we typically assume the photosphere to have receded into the inner layers of the SN, and for the ejecta to have expanded, cooled, and become optically thin, reaching a low enough density for these forbidden transition emission lines to occur. This therefore often requires the SN to be deep in its nebular phase for these features to be observed.

Across our sample of Ca-strong SNe, however, we detect these forbidden emission lines from as early as peak light -- far earlier than this typical understanding predicts. To explain the early presence of these features, we investigated the behaviour of the forbidden lines extensively, and drew the following conclusions:

\begin{itemize}
    \item The early presence of [Ca~II] and [O~I] emission in each object are consistent with the presence of external low-density Ca and O regions that pre-date the SN. 
    \item The broad, non-zero, distribution of velocities in the photospheric phase suggests an extended aspherical external system. Such a system would reproduce a broad range of measured velocities dependant upon the observed angle of the SN, in line with our observations.
    \item The narrower, zero-centred, velocity distribution of these forbidden lines in the truly nebular-phase spectra of each object may originate from the SN core. 
    \item Investigating higher-resolution spectra observed during transitional epochs revealed a complex line evolution: an initial high-velocity component which is gradually dominated by a zero-velocity component at later times. This transition is consistent with a shift in the dominant emission region from the external aspherical material at early times, to the inner core at late times as the ejecta expand, become optically thin, and the photosphere recedes entirely.
    \item Analysis focussed on the early flux excesses observed before peak light in the majority of our sample of Ca-strong SNe indicates that these objects occur in environments that are polluted by pre-existing, extended material.
\end{itemize}

Our results are consistent with the contribution from two distinct emission regions in the structure of Ca-strong SNe: an external, aspherical structure responsible for early-time forbidden line emission -- interaction with which may be responsible for the observed early photometric behaviour of Ca-strong SNe -- and a central, core component that dominates in the nebular phase. We therefore hypothesise that Ca-strong SNe result from WD explosions in polluted environments. 

We suggest this pollution may originate from precursor activity of the progenitor itself, e.g. through prior He-novae in AM CVn-based systems. Following our previous work in \citetalias{2025MNRAS.537.1015T}, we suggest that the observed behaviours across the sample of Ca-strong SNe are consistent with the proposed system of a Ca-strong SN progenitor embedded in an environment polluted by an AM CVn recurrent He-nova. In such a scenario, the polar outflow lobes produced by prior nova activity are analogous to the external aspherical structure we conclude is responsible for the photospheric [Ca~II] and [O~I] emission. The observed emergence of nebular emission would instead be directly produced from the core of the SN itself.

\begin{acknowledgements}

The full version of all Tables presented are available in the online supplementary material. 

C.-G.T.-P., K.M., J.H.T. acknowledge funding from Horizon Europe ERC grant no.~101125877.
    
W.J.-G.\ is supported by NASA through Hubble Fellowship grant HSTHF2-51558.001. 

MP acknowledges support from a UK Research and Innovation Future Leaders Fellowship (grant references MR/T020784/1 and UKRI1062)

CRA is supported by the European Research Council (ERC) under the European Union's Horizon 2020 research and innovation program (grant agreement No. 948381, PI: M. Nicholl).

C.O.D acknowledges funding from the Irish Research Council grant no GOIPG/2024/4378.

C.L. is supported by DoE award \#\,DE-SC0025599.

A.C.G. and the Fong Group at Northwestern acknowledge support by the National Science Foundation under grant Nos. AST-1909358, AST-2206494, AST-2308182 and CAREER grant No. AST-2047919.

SJS acknowledges funding from STFC Grants ST/Y001605/1, ST/X001253/1, a Royal Society Research Professorship and the Hintze Family Charitable Foundation. 

Based on observations obtained with the Samuel Oschin Telescope 48-inch and the 60-inch Telescope at the Palomar Observatory as part of the Zwicky Transient Facility project. ZTF is supported by the National Science Foundation under Grants No. AST-1440341 and AST-2034437 and a collaboration including current partners Caltech, IPAC, the Oskar Klein Center at Stockholm University, the University of Maryland, University of California, Berkeley , the University of Wisconsin at Milwaukee, University of Warwick, Ruhr University Bochum, Cornell University, Northwestern University and Drexel University. Operations are conducted by COO, IPAC, and UW.

SED Machine is based upon work supported by the National Science Foundation under Grant No. 1106171.

The ZTF forced-photometry service was funded under the Heising-Simons Foundation grant 12540303 (PI: Graham).
    
The ztfquery code was funded by ERC grant no. 759194 - USNAC, PI: Rigault).

W. M. Keck Observatory access was supported by Northwestern University and the Center for Interdisciplinary Exploration and Research in Astrophysics (CIERA).

Pan-STARRS is primarily funded to search for near-Earth asteroids through NASA grants NNX08AR22G and NNX14AM74G. The Pan-STARRS science products were made possible through the contributions of the University of Hawai’i Institute for Astronomy, the Queen’s University Belfast and the University of Oxford. 

This work has made use of data from the Asteroid Terrestrial-impact Last Alert System (ATLAS) project. The Asteroid Terrestrial-impact Last Alert System (ATLAS) project is primarily funded to search for near earth asteroids through NASA grants NN12AR55G, 80NSSC18K0284, and 80NSSC18K1575; byproducts of the NEO search include images and catalogs from the survey area. This work was partially funded by Kepler/K2 grant J1944/80NSSC19K0112 and HST GO-15889, and STFC grants ST/T000198/1 and ST/S006109/1. The ATLAS science products have been made possible through the contributions of the University of Hawaii Institute for Astronomy, the Queen’s University Belfast, the Space Telescope Science Institute, the South African Astronomical Observatory, and The Millennium Institute of Astrophysics (MAS), Chile and the University of Oxford.

\end{acknowledgements}

\bibliographystyle{aa} 
\bibliography{ca_rich}

\appendix

\section{Reclassifications of literature transients}
\label{reclassify}

\cite{2023ApJ...959...12D} presented a sample of `IIb-Ca-strong' objects -- objects with a high [Ca~II]/[O~I] ratio at late-times, but with broad H features in the photospheric spectra akin to Type IIb SNe. Our analysis of the photospheric spectra of these objects, however, indicates the features previously identified as H in several of these objects would require unphysical velocities, with the absorption occurring at 0~km~s\textsuperscript{-1}. The velocities of these features among this subsample of `IIb-Ca-strong' objects are consistent with the expected velocities of He I at these phases. Additionally, if this were He I, the required velocity for this observed feature is consistent across each feature in the He~I~\(\lambda\)4471,~\(\lambda\)5876,~\(\lambda\)6678,~\(\lambda\)7065 series, further indicating the chemical species responsible for this feature to be He I. Due to this, we update the classification of SN~2019ehk, SN~2019pof, and SN~2021sjt \citep{ayala2025early, jacobson2020sn, nakaoka2021calcium, 2021TNSCR2383....1S, 2019TNSCR2097....1D, 2023ApJ...959...12D, 2019TNSCR.675....1D}  and include them in our sample of `confirmed' Ca-strong objects. 

We reclassify five candidate Ca-strong SNe from \cite{de2020zwicky} for the following reasons:

\emph{SN~2019txr} (SN~Ib/c): the +270d spectrum, from which the \cite{de2020zwicky} [Ca~II]/[O~I] ratio was determined, is host-dominated and has a very low signal-to-noise; we did not clearly identify any distinct features in this spectrum. We instead obtain our [Ca~II]/[O~I] value from the +30d spectrum ([Ca~II]/[O~I]$=>2.21$), a phase consistent with our definition of the nebular phase. As this value meets Criterion 4 of our classification requirements, we include SN~2019txr as a Ca-strong object.

\emph{SN~2019ouq} (SN Ic): Fitting the [Ca~II] and [O~I] features of the +180d spectrum with the methodology described in \ref{speccy time}, we obtain a [Ca~II]/[O~I] ratio of 3.62. As this is above the threshold value of 2 (see Criterion 4), we therefore include SN~2019ouq in our sample. 

\emph{SN~2018kqr} (SN Ic-BL): The spectrum at +14d is still in the transitional-phase between the photospheric and nebular regimes. \cite{de2020zwicky} describe this object as having slow photospheric-phase evolution, but without later spectra (or a better-constrained value on its photometric decline rate, $\Delta m_7$; see Section \ref{lighty time}) we cannot comfortably come to the same conclusion. We therefore continue to treat this object as a Ca-strong SN as it satisfies each of our classification criteria.

\emph{SN~2019gau} (SN Ia): While the only late-time spectrum of this object (+250d) is contaminated by host emission, we tentatively identify a [Ca II] feature. As this spectrum is so dominated by host emission, and we cannot concretely confirm the presence of [Ca~II], we have treated this object as though its latest-available spectrum was the previous spectrum, at +9d, and proceeded with its classification under this assumption. Its spectrum at this epoch resembled that of other confirmed Ca-strong objects, and its photometric properties are in line with the rest of the sample, so we continued to assume SN~2019gau to be a Ca-strong object.

\emph{SN~2018dbg} (SN Ib/c): The only spectrum available (+28d) contains distinct NIR triplet, [Ca II], and [O I] emission. Additionally, the photometic behaviour (M\textsubscript{peak}, $\Delta m_7$, $g-r$ colour) of this object was in line with the rest of the sample. We therefore proceeded with our classification of SN~2018dbg as a member of our full Ca-strong sample.

\section{Additional plots and  tables}

\begin{figure*}[!htbp]
    \centering
	\includegraphics[width=\textwidth]{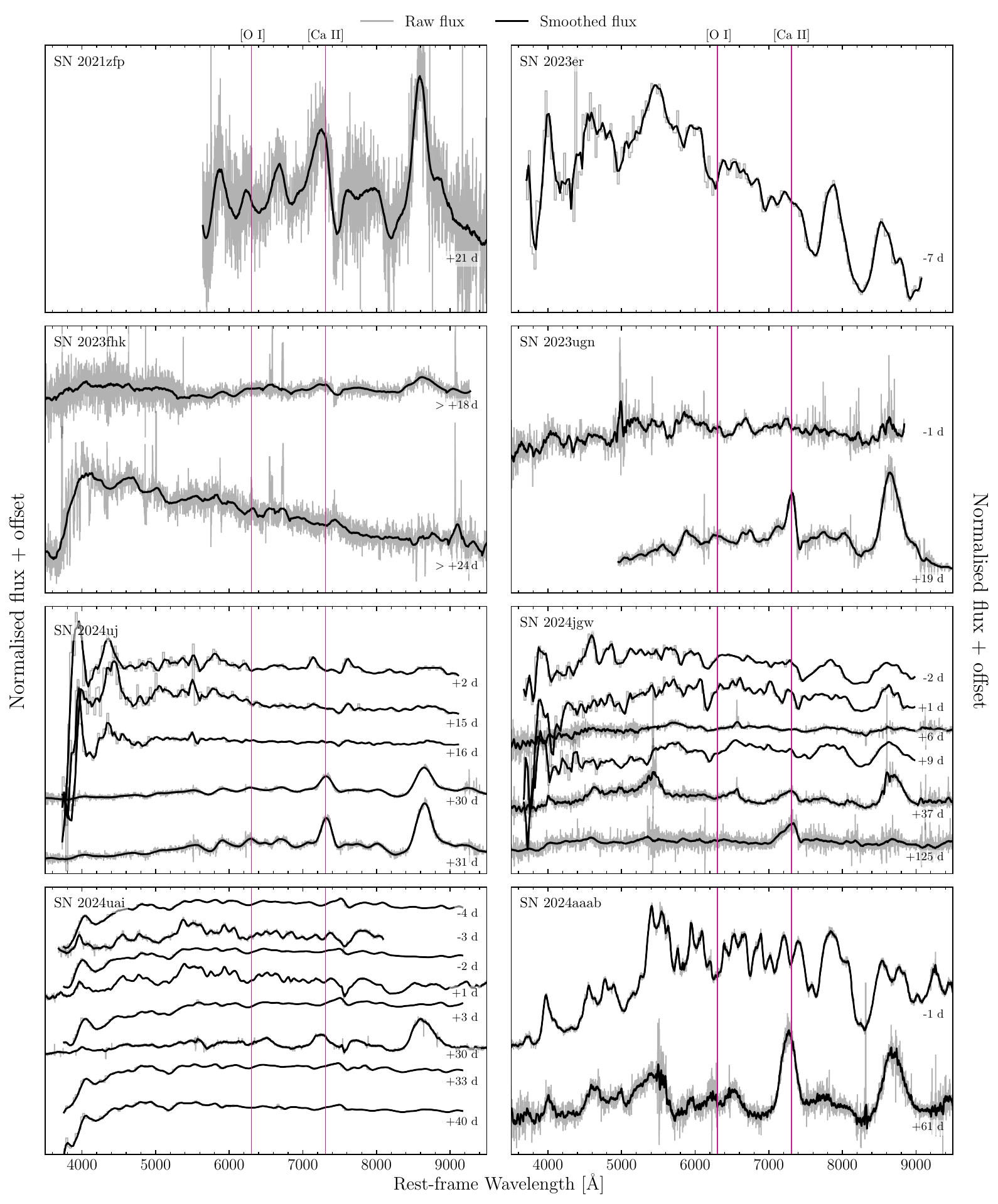}
    \caption{Spectral time series for each of the 8 confirmed new Ca-strong SNe in our sample. Each spectrum (light grey)
    is overlaid with a Savitzky-Golay smoothed spectrum (black). The rest-frame phase relative to $r-$band peak of each spectroscopic observation is labelled. Any narrow host emission lines present in the spectra, namely for objects SN~2022psx and SN~2023fhk, have been masked from the spectra to ensure consistent scaling of the SN spectral emission between epochs and across objects. Pink vertical lines correspond to the weighted mean wavelength values of the [O~I]~\(\lambda\lambda\)6300,6364 and [Ca~II]~\(\lambda\lambda\)7291,7324 features respectively, based on their relative component line intensities.}
    \label{fig:new_spectra}
\end{figure*}

\begin{figure*}[!htbp]
	\includegraphics[width=\textwidth]{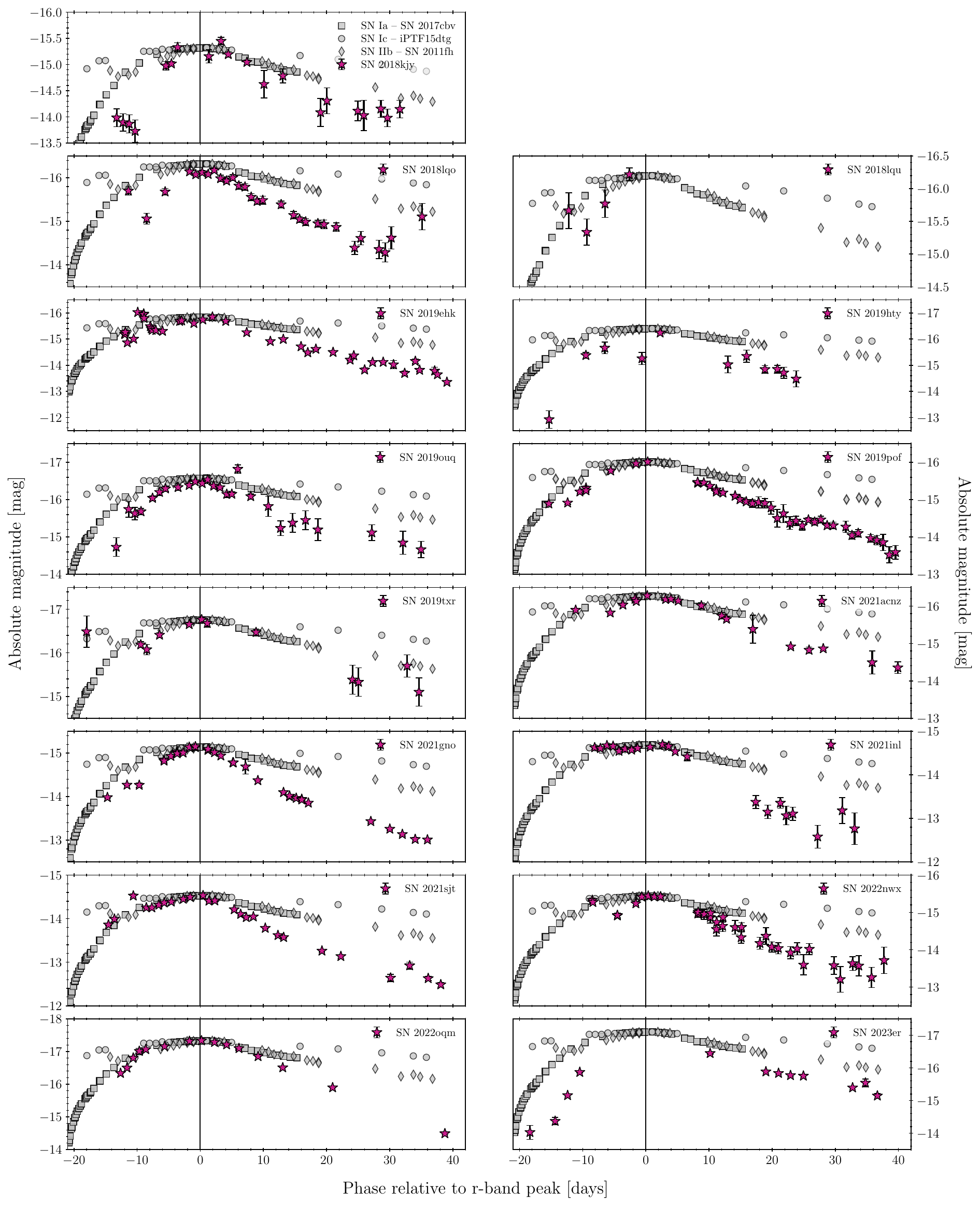}
    \caption{$r-$band light curves of the 15 objects in our sample with confirmed early flux excesses. Each object (labelled in subplot; left to right, top to bottom: SN~2018kjy, SN~2018lqo, SN~2018lqu, SN~2019ehk, SN~2019hty, SN~2019ouq, SN~2019pof, SN~2019txr, SN~2021acnz, SN~2021gno, SN~2021inl, SN~2021sjt, SN~2022nwx, SN~2022oqm, SN~2023er) is plotted against the light curves of known SNe with PPBs -- SN~Ia (SN~2017cbv, squares), SN~Ic (iPTF15dtg, circles), and SN~IIb (SN~2011fh, diamonds) -- scaled to the peak brightness of that object.}
    \label{fig:bumps_iib_comparison}
\end{figure*}

\nolinenumbers
\clearpage
\onecolumn
\begin{landscape}
\centering

\begin{longtable}{l l l l c c c c l}
\caption{Full sample of all Ca-strong SN candidates. IAU name, ZTF name, host galaxy, redshift, extinction magnitude, projected offset from host galaxy, [Ca~II]/[OI] ratio, and associated references are presented. Confirmed and rejected Ca-strong objects from the full sample of candidates are labelled accordingly. Objects SN~2019gsc \citep{2020ApJ...892L..24S} and SN~2019ttf have been previously classified as SNe Iax and exhibit strong matches with this SN class around peak; we therefore exclude them from our sample. Objects with redshift values marked with * had no uncertainty available. The redshift of SN~2023ugn -- denoted with $\ddagger$ -- has been separately reported as two different values: $0.03224 \pm 0.00008$ and $0.014673 \pm 0.000067$. Based on the observed wavelength of the Ca~II NIR triplet in the latest spectrum of this object, we use a redshift value of $0.03224 \pm 0.00008$. Object [Ca~II]/[O~I] values marked with $\dagger$ had no spectroscopic observations after +30~d, and as such their nebular [Ca~II]/[O~I] ratio could not be measured; objects with upper limits of this value presented had [O~I] emission of SNR<3 in their latest available epoch.}
\label{tab:full_sample} \\

\hline
& IAU Name & ZTF Name & Host & z & E(B-V) [Mag] & R\textsubscript{off} [kpc] & [Ca~II]/[O~I] & Reference\\
\hline
\endfirsthead

\multicolumn{9}{c}%
{\tablename\ \thetable\ -- \textit{Continued from previous page}} \\
\hline
& IAU Name & ZTF Name & Host & z & E(B-V) [Mag] & R\textsubscript{off} [kpc] & [Ca~II]/[O~I] & Reference\\
\hline
\endhead

\hline \multicolumn{9}{r}{\textit{Continued on next page}} \\
\endfoot

\hline
\endlastfoot

\multirow{46}{*}{\rotatebox{90}{Confirmed Ca-strong}} & SN 2000ds &  & NGC 2768 & 0.004513 $\pm$ 0.0000170 & 0.0374 $\pm$ 0.0020 & 3.54 $\pm$ 0.39 & $\dagger$ & {\footnotesize 1, 2, 3} \\
& SN 2001co &  & NGC 5559 & 0.017175 $\pm$ 0.00000973 & 0.0169 $\pm$ 0.0002 & 6.91 $\pm$ 0.20 & $\geq2.51$ & {\footnotesize 4, 2, 3} \\
& SN 2003H &  & NGC 2207 & 0.009176 $\pm$ 0.00000200 & 0.0748 $\pm$ 0.0006 & 10.59 $\pm$ 0.57 & 4.48 $\pm$ 0.12 & {\footnotesize 5, 2, 3} \\
& SN 2003dg &  & UGC 6934 & 0.018339 $\pm$ 0.00000700 & 0.0205 $\pm$ 0.0016 & 1.22 $\pm$ 0.03 & 7.89 $\pm$ 0.09 & {\footnotesize 6, 2, 3} \\
& SN 2003dr &  & NGC 5714 & 0.007472 $\pm$ 0.00000500 & 0.0125 $\pm$ 0.0009 & 2.46 $\pm$ 0.16 & 7.62 $\pm$ 0.14 & {\footnotesize 5, 2, 3} \\
& SN 2005E &  & NGC 1032 & 0.009150 $\pm$ 0.0000120 & 0.0315  $\pm$ 0.0012 & 26.37 $\pm$ 1.47 & 6.67 $\pm$ 0.09 &  {\footnotesize 5, 7, 8, 9, 3} \\
& SN 2005cz &  & NGC 4589 & 0.006605 $\pm$ 0.0000467 & 0.0241 $\pm$ 0.0006 & 6.06 $\pm$ 0.45 & 11.20 $\pm$ 0.08 & {\footnotesize 10, 8, 11, 2, 3} \\
& SN 2007ke &  & NGC 1129 & 0.017472 $\pm$ 0.0000470 & 0.0949 $\pm$ 0.0039 & 22.28 $\pm$ 0.63 & $\dagger$ & {\footnotesize 12, 2, 3} \\
& PTF09dav &  & 2MASX J22465295+2138221 & 0.037100 $\pm$ 0.000200 & 0.0398 $\pm$ 0.0005 & 45.05 $\pm$ 0.61 & $\geq4.37$ & {\footnotesize 13, 13, 14, 2, 3} \\
& SN 2010et &  & SDSS J171650.20+313234.4 & 0.023461 $\pm$ 0.0000132 & 0.0366 $\pm$ 0.0009 & 48.21 $\pm$ 1.03 & 14.83 $\pm$ 0.03 & {\footnotesize 4, 15, 2, 3} \\
& PTF11kmb &  & NGC 7265 & 0.016661 $\pm$ 0.0000477 & 0.0905 $\pm$ 0.0021 & 174.08 $\pm$ 5.15 & 10.18 $\pm$ 0.10 & {\footnotesize 12, 2, 16, 3} \\
& PTF12bho &  & LEDA 44863 & 0.023430 $\pm$ 0.0000100 & 0.0080 $\pm$ 0.0004 & 56.45 $\pm$ 1.20 & 12.37 $\pm$ 0.07 & {\footnotesize 17, 18, 16, 3} \\
& PTF15eqv &  & NGC 3430 & 0.005290 $\pm$ 0.00000300 & 0.0201 $\pm$ 0.0002 & 4.89 $\pm$ 0.53 & 5.07 $\pm$ 0.07 & {\footnotesize 19, 20, 3, 21} \\
& SN 2016hgs &  & 2MASX J00505254+2722432 & 0.017000 & 0.0528 $\pm$ 0.0018 & 6.38 $\pm$ 0.19 & 7.09 $\pm$ 0.07 &  {\footnotesize 22, 3, 23} \\
& SN 2016hnk &  & MCG-01-06-070 & 0.015942 $\pm$ 0.00000700 & 0.0224 $\pm$ 0.0008 & 4.23 $\pm$ 0.13 & $\geq6.82$ &  {\footnotesize 24, 25, 3, 26, 27} \\
& SN 2018ckd & ZTF18aayhylv & NGC 5463 & 0.023649 $\pm$ 0.0000477 & 0.0243 $\pm$ 0.0004 & 19.55 $\pm$ 0.41 & 4.16 $\pm$ 0.02 & {\footnotesize 12, 3} \\
& SN 2018dbg & ZTF18abdffeo & IC 4397 & 0.014744 $\pm$ 0.0000200 & 0.0174 $\pm$ 0.0004 & 0.65 $\pm$ 0.02 & $\dagger$ & {\footnotesize 28, 3} \\
& SN 2018gwo & ZTF18acbwazl & NGC 4128 & 0.007442 $\pm$ 0.000051 & 0.0187 $\pm$ 0.0006 & 16.46 $\pm$ 0.01 & 5.65 $\pm$ 0.07 & {\footnotesize 12, 3} \\
& SN 2018kjy & ZTF18acsodbf & NGC 2256 & 0.017519 $\pm$ 0.000193 & 0.1125 $\pm$ 0.0016 & 28.89 $\pm$ 0.94 & 2.97 $\pm$ 0.62 & {\footnotesize 6, 3} \\
& SN 2018kqr & ZTF18acushie & SDSS J085003.97+551010.1 & 0.04425 $\pm$ 0.00001 & 0.0293 $\pm$ 0.0016 & 5.07 $\pm$ 0.06 & $\dagger$ & {\footnotesize 29, 3} \\
& SN 2018lqo & ZTF18abmxelh & Z 224-43 & 0.032280 $\pm$ 0.00000900 & 0.0059 $\pm$ 0.0004 & 18.91 $\pm$ 0.29 & 12.19 $\pm$ 0.08 & {\footnotesize 24, 3} \\
& SN 2018lqu & ZTF18abttsrb & WISEA J155413.91+133102.4 & 0.034826 $\pm$ 0.0000231 & 0.0359 $\pm$ 0.0007 & 28.42 $\pm$ 0.40 & 3.93 $\pm$ 0.06 & {\footnotesize 4, 3} \\
& SN 2019bkc &  & NGC 3090 & 0.020311 $\pm$ 0.0000670 & 0.0453 $\pm$ 0.0035 & 91.97 $\pm$ 2.26 & $\dagger$ & {\footnotesize 30, 31, 32, 33} \\
& SN 2019ehk & ZTF19aatesgp & M 100 & 0.005240 $\pm$ 0.00000300 & 0.0227 $\pm$ 0.0002 & 2.44 $\pm$ 0.23 & $\geq3.42$ & {\footnotesize 34, 35, 36, 37, 38} \\
& SN 2019gau & ZTF19aavlfvn & 2MASS J14381034+1008062 & 0.027683 $\pm$ 0.0000131 & 0.0261 $\pm$ 0.0006 & 0.80 $\pm$ 0.01 & $\dagger$ &  {\footnotesize 4, 3} \\
& SN 2019hty & ZTF19aaznwze & WISEA J125534.50+321221.5 & 0.022829 $\pm$ 0.00000676 & 0.0141 $\pm$ 0.0008 & 9.20 $\pm$ 0.19 & 2.33 $\pm$ 0.03 & {\footnotesize 39, 3} \\
& SN 2019ofm & ZTF19abrdxbh & IC 4514 & 0.030254 $\pm$ 0.0000106 & 0.0248 $\pm$ 0.0014 & 11.96 $\pm$ 0.20 & 4.89 $\pm$ 0.04 & {\footnotesize 4, 3} \\
& SN 2019ouq & ZTF19abhhdwf & LEDA 1890714 & 0.035864 $\pm$ 0.0000247 & 0.0259 $\pm$ 0.0004 & 7.71 $\pm$ 0.11 & 3.62 $\pm$ 0.02 & {\footnotesize 4, 3} \\
& SN 2019pof & ZTF19abxtcio & PSO J018.1606+33.0353 & 0.0155 & 0.0508 $\pm$ 0.0008 & 3.21 $\pm$ 0.11 & $\geq5.21$ & {\footnotesize 40, 37} \\
& SN 2019pxu & ZTF19abwtqsk & LEDA 147063 & 0.028200 $\pm$ 0.000150 & 0.1026 $\pm$ 0.0035 & 18.13 $\pm$ 0.44 & 4.29 $\pm$ 0.32 &  {\footnotesize 41, 3} \\
& SN 2019txr & ZTF19aarrdoz & LEDA 2532103 & 0.044433 & 0.0337 $\pm$ 0.0026 & 3.26 $\pm$ 0.04 & $\geq2.21$ & {\footnotesize 42, 3} \\
& SN 2021gno & ZTF21aaqhhfu & NGC 4165 & 0.006231 $\pm$ 0.0000157 & 0.0295 $\pm$ 0.0013 & 3.21 $\pm$ 0.26 & $\dagger$ & {\footnotesize 4, 43, 3} \\
& SN 2021inl & ZTF21aasuego & NGC 4923 & 0.018257 $\pm$ 0.00000659 & 0.0077 $\pm$ 0.0004 & 22.79 $\pm$ 0.63 & $\dagger$ & {\footnotesize 44, 43} \\
& SN 2021sjt & ZTF21abjyiiw & NGC 6951 & 0.004750 $\pm$ 0.00000334 & 0.3178 $\pm$ 0.0086 & 7.54 $\pm$ 0.79 & $\dagger$ & {\footnotesize 45, 37} \\
& AT 2021zfp & ZTF21acddxti & 2MASX J10133383+3842365 & 0.023034 $\pm$ 0.00000700 & 0.0111 $\pm$ 0.0002 & 13.12 $\pm$ 0.28 & $\dagger$ & {\footnotesize 46, 47} \\
& SN 2021acnz & ZTF21aciwixf & NGC 165 & 0.019644 $\pm$ 0.00000667 & 0.0296 $\pm$ 0.0002 & 40.38 $\pm$ 1.11 & 7.33 $\pm$ 0.01 & {\footnotesize 48, 49} \\
& SN 2022nwx & ZTF22aapisdk & NGC 7242 & 0.019304 $\pm$ 0.0000407 & 0.1289 $\pm$ 0.0021 & 37.87 $\pm$ 0.99 & 7.81 $\pm$ 0.05 & {\footnotesize 12} \\
& SN 2022oqm & ZTF22aasxgjp & NGC 5875 & 0.011695 $\pm$ 0.000007 & 0.0164 $\pm$ 0.0005 & 18.44 $\pm$ 0.82 & 6.90 $\pm$ 0.04 & {\footnotesize 50, 51} \\
& SN 2023er & ZTF22abyrurb & 2MASX J01021072+3224501 & 0.016285 $\pm$ 0.0000870 & 0.0653 $\pm$ 0.0010 & 123.15 $\pm$ 3.68 & $\dagger$ & {\footnotesize 52, 53, 54} \\
& SN 2023fhk & ZTF23aaawmyi & 2MASX J22301302+3703304 & 0.024200 & 0.1067 $\pm$ 0.0019 & 0.57 $\pm$ 0.23 & $\dagger$ & {\footnotesize 55, 56, 57} \\
& SN 2023ugn &  & NGC 915 & 0.03224 $\pm$ 0.00008 $\ddagger$ & 0.0894 $\pm$ 0.0047 & 13.02 $\pm$ 0.22 & $\dagger$ & {\footnotesize 6, 58, 59} \\
& SN 2023xwi & ZTF23abrgldj & PSO J146.8867+71.243 & 0.0112 $\pm$ 0.0005 & 0.0557 $\pm$ 0.0026 & 9.12 $\pm$ 0.41 & 23.02 $\pm$ 0.02 & {\footnotesize 60} \\
& SN 2024uj & ZTF24aabuxza & NGC 3565 & 0.012612 $\pm$ 0.000130 & 0.0423 $\pm$ 0.0024 & 5.93 $\pm$ 0.24 & 4.57 $\pm$ 0.07 & {\footnotesize 61, 62, 63} \\
& SN 2024jgw & ZTF24aaosvzx & WISEA J171503.17+460243.5 & 0.028 * & 0.0302 $\pm$ 0.0009 & 2.86 $\pm$ 0.05 & $\dagger$ & {\footnotesize 64, 65} \\
& SN 2024uai & ZTF24abejmqz & NGC 2300 & 0.006354 $\pm$ 0.0000230 & 0.0815 $\pm$ 0.0043 & 3.45 $\pm$ 0.29 & 5.07 $\pm$ 0.09 & {\footnotesize 66, 67, 68} \\
& AT 2024aaab & ZTF24abpgriz & Z 307-16 & 0.017359 $\pm$ 0.000160 & 0.2296 $\pm$ 0.0123 & 19.04 $\pm$ 0.55 & $\geq3.76$ & {\footnotesize 6, 69} \\
\hline
        \multirow{20}{*}{\rotatebox{90}{Not Ca-strong}} & SN 2012hn &  & NGC 2272 & 0.007104 $\pm$ 0.0000320 & 0.0833 $\pm$ 0.0039 & -- & 1.63 $\pm$ 0.19 & {\footnotesize 70, 71, 2, 3} \\
& SN 2018fob & ZTF18aboabxv & NGC 5888 & 0.029050 $\pm$ 0.0000110 & 0.0178 $\pm$ 0.0020 & -- & 0.49 $\pm$ 0.03 & {\footnotesize 4, 3} \\
& SN 2019ttf & ZTF19abalbim & Z 143-17 & 0.011221 $\pm$ 0.0000113 & 0.1276 $\pm$ 0.0031 & 2.81 $\pm$ 0.12 & $\dagger$ & {\footnotesize 72, 3} \\
& SN 2019gsc & ZTF19aawhlcn & Z 273-16 & 0.011299 $\pm$ 0.0000113 & 0.0081 $\pm$ 0.0003 & 4.10 $\pm$ 0.18 & 2.51 $\pm$ 0.02 & {\footnotesize 4, 3} \\
& SN 2018gjx & ZTF18abwkrbl & NGC 865 & 0.009990 $\pm$ 0.00000500 & 0.0758 $\pm$ 0.0009 & -- & 3.01 $\pm$ 0.09 & {\footnotesize 5, 73, 3, 37} \\
& SN 2018jak & ZTF18acqxyiq & LEDA 2056276 & 0.038491 $\pm$ 0.0000102 & 0.0105 $\pm$ 0.0004 & -- & 1.01 $\pm$ 0.04 & {\footnotesize 4, 3, 37} \\
& SN 2019yz & ZTF19aadttht & UGC 9977 & 0.006388 $\pm$ 0.00000900 & 0.0949 $\pm$ 0.0022 & -- & 0.50 $\pm$ 0.02 & {\footnotesize 5, 3} \\
& SN 2019abb & ZTF19aadwtoe & Z 58-66 & 0.015295 $\pm$ 0.00000600 & 0.0323 $\pm$ 0.0016 & -- & 0.41 $\pm$ 0.04 & {\footnotesize 74, 3} \\
& SN 2019ape & ZTF19aailcgs & NGC 3426 & 0.020305 $\pm$ 0.00000751 & 0.0302 $\pm$ 0.0006 & -- & 1.29 $\pm$ 0.03 & {\footnotesize 4, 3} \\
& SN 2019ccm & ZTF19aamfupk & UGC 3110 & 0.014947 $\pm$ 0.0000170 & 0.140 $\pm$ 0.006 & -- & 1.05 $\pm$ 0.01 & {\footnotesize 5} \\
& SN 2019mjo & ZTF19abgqruu & MCG+00-01-026 & 0.042196 * & 0.018 $\pm$ 0.007 & -- & 1.42 $\pm$ 0.04 & {\footnotesize 75, 3} \\
& SN 2019txl & ZTF19aanfsmc & UGC 5084 & 0.033806 $\pm$ 0.00000969 & 0.018 $\pm$ 0.008 & -- & 0.54 $\pm$ 0.01 &  	{\footnotesize 4, 3} \\
& SN 2019txt & ZTF19aasqseq & IC 583 & 0.026391 $\pm$ 0.00000886 & 0.029 $\pm$ 0.009 & -- & 0.39 $\pm$ 0.04 & {\footnotesize 4, 3} \\
& SN 2019weu & ZTF19acxgxcu & WISEA J232432.49+050638.5 & 0.038 * & 0.048 $\pm$ 0.007 & -- & $\dagger$ & {\footnotesize 3} \\
& SN 2020sbw & ZTF20abwzqzo & CGCG 389-007 & 0.022769 $\pm$ 0.00000900 & 0.0325 $\pm$ 0.0007 & -- & 8.06 $\pm$ 0.07 &  {\footnotesize 5, 37} \\
& SN 2021M &  & UGC 9113 & 0.010624 $\pm$ 0.0000149 & 0.0197 $\pm$ 0.0012 & -- & 6.87 $\pm$ 0.07 & {\footnotesize 4, 37} \\
& SN 2021pb & ZTF21aabxjqr & 2MASX J09444707+5141192 & 0.033178 $\pm$ 0.0000117 & 0.0088 $\pm$ 0.0009 & -- & 1.27 $\pm$ 0.15 & {\footnotesize 4, 37} \\
& AT 2021niv & ZTF21abmkuji & LEDA 1102966 & 0.018419 $\pm$ 0.000150 & 0.032 $\pm$ 0.005 & -- & $\dagger$ & {\footnotesize 41, 3} \\
& SN 2022psx & ZTF22aawcmcr & WISEA J230439.90+372359.9 & 0.019 * & 0.137 $\pm$ 0.002 & -- & 7.51 $\pm$ 0.03 & {\footnotesize 76} \\
& SN 2022sfh & ZTF22abdgnuc & 2MASX J02365426+3752234 & 0.032452 $\pm$ 0.0000930 & 0.045 $\pm$ 0.004 & -- & 13.28 $\pm$ 0.09 & {\footnotesize 77} 

\end{longtable}

\begin{multicols}{5}
\footnotesize
    \begin{enumerate}
    \item \citet{2011MNRAS.413..813C}
    \item \citet{foley2015kinematics}
    \item \citet{de2020zwicky}
    \item \citet{2017ApJS..233...25A}
    \item \citet{2005ApJS..160..149S}
    \item \citet{1999PASP..111..438F}
    \item \citet{2010Natur.465..322P}
    \item \citet{2010Natur.465..326K}
    \item \citet{2011ApJ...738...21W}
    \item \citet{2019ApJ...871...33L}
    \item \citet{2011ApJ...728L..36P}
    \item \citet{2015ApJS..218...10V}
    \item \citet{2011ApJ...732..118S}
    \item \citet{kasliwal2012calcium}
    \item \citet{2010CBET.2339....1D}
    \item \citet{lunnan2017two}
    \item \citet{2021AA...650A..76H}
    \item \citet{2017yCat..18360060L}
    \item \citet{1997AJ....114...77N}
    \item \citet{milisavljevic2017iptf15eqv}
    \item \citet{2018AAS...23144603K}
    \item \citet{2022ApJ...927..199D}
    \item \citet{de2018iptf}
    \item \citet{2017ApJ...835..280L}
    \item \citet{galbany2019evidence}
    \item \citet{2018MNRAS.475L.111S}
    \item \citet{2020ApJ...896..165J}
    \item \citet{2003AA...412...57P}
    \item \citet{2015ApJS..219...12A}
    \item \citet{2019TNSTR.310....1T}
    \item \citet{2019TNSCR.336....1P}
    \item \citet{1995A&A...300..675H}
    \item \citet{prentice2020rise}
    \item \citet{2014MNRAS.440..696A}
    \item \citet{jacobson2020sn}
    \item \citet{nakaoka2021calcium}
    \item \citet{2023ApJ...959...12D}
    \item \citet{2021ApJ...907L..18D}
    \item \citet{2016ApJ...819...63R}
    \item \citet{ayala2025early}
    \item \citet{2009MNRAS.399..683J}
    \item \citet{2018MNRAS.474.1873W}
    \item \citet{jacobson2022circumstellar}
    \item \citet{2024AJ....168...58D}
    \item \citet{2024ApJ...964..172B}
    \item \citet{2006ApJS..162...38A}
    \item \citet{2021TNSTR3269....1D}
    \item \citet{2016AA...595A.118V}
    \item Frohmaier+ in prep.
    \item \citet{2016A&A...595A.118V}
    \item \citet{2024ApJ...962..109I}
    \item \citet{1994AJ....108...33S}
    \item \citet{2023TNSTR..57....1H}
    \item \citet{2024TNSCR.684....1S}
    \item \citet{1998AJ....116.1094C}
    \item \citet{2023TNSTR.776....1D}
    \item \citet{2024TNSCR3553....1D}
    \item \citet{2023TNSTR2518....1C}
    \item \citet{2023TNSCR2698....1F}
    \item \citet{2025MNRAS.537.1015T}
    \item \citet{1998AJ....116....1D}
    \item \citet{2024TNSTR..85....1T}
    \item \citet{2024TNSCR.110....1P}
    \item \citet{2024TNSTR1638....1F}
    \item \citet{2024TNSCR3029....1J}
    \item \citet{2012MNRAS.426..296D}
    \item \citet{2024TNSTR3213....1R}
    \item \citet{2024TNSCR3257....1R}
    \item \citet{2024TNSTR4282....1S}
    \item \citet{2006MNRAS.371.1855W}
    \item \citet{2015MNRAS.450.4198S}
    \item \citet{2019AJ....158..234S}
    \item \citet{2020MNRAS.499.1450P}
    \item \citet{2008ApJS..175..297A}
    \item \citet{2014ApJS..210....9B}
    \item \citet{2022TNSCR2170....1H}
    \item \citet{1999ApJS..121..287H}

\end{enumerate}
\end{multicols}

\end{landscape}
\twocolumn

\begin{table*}
    \centering
    \caption{Optical spectroscopy of the 8 newly-confirmed Ca-strong objects in our sample.}
    \begin{tabular}{lcccc}
    \hline
    Object & MJD & Phase [days]  & Telescope & Instrument \\
    \hline
    SN 2021zfp & 59482 & +3 & Keck II & NIRES \\
               & 59500 & +21 & P200 & DBSP \\
    SN 2023er  & 59940 & $-$7 & P60 & SEDm$^{*}$ \\
    SN 2023fhk & 60054 & $\geq$+18 & P200 & DBSP \\
               & 60060 & $\geq$+24 & Keck I & LRIS \\
    SN 2023ugn & 60231 & $-$1 & UH88 & SNIFS \\
               & 60251 & +19 & -- & -- \\
    SN 2024uj  & 60326 & +2 & P60 & SEDm$^{*}$ \\
               & 60339 & +15 & P60 & SEDm$^{*}$ \\
               & 60340 & +16 & P60 & SEDm$^{*}$ \\
               & 60354 & +30 & Keck I & LRIS \\
               & 60355 & +31 & P60 & SEDm$^{*}$ \\
    SN 2024jgw & 60461 & -2 & P60 & SEDm$^{*}$ \\
               & 60464 & +1 & P60 & SEDm$^{*}$ \\
               & 60468 & +6 & Lick & KAST \\
               & 60472 & +9 & P60 & SEDm$^{*}$ \\
               & 60500 & +37 & P200 & DBSP \\
               & 60588 & +125 & Keck I & LRIS \\
    SN 2024uai & 60554 & -4 & P60 & SEDm$^{*}$ \\
               & 60555 & -3 & Copernico 1.82m & AFOSC \\
               & 60556 & -2 & P60 & SEDm$^{*}$ \\
               & 60559 & +1 & NOT & ALFOSC \\
               & 60561 & +3 & P60 & SEDm$^{*}$ \\
               & 60588 & +30 & Keck I & LRIS \\
               & 60591 & +33 & P60 & SEDm$^{*}$ \\
               & 60598 & +40 & P60 & SEDm$^{*}$ \\
    SN 2024aaab & 60380 & $-$1 & Keck I & LRIS \\
                & 60707 & +61 & Keck I & LRIS \\
    \hline
    \end{tabular}
    
    \vspace{2mm}
    \raggedright
    $^{*}$SEDm spectra obtained with the Spectral Energy Distribution Machine \citep{2019A&A...627A.115R}.
    
    \label{tab:spec_info}
\end{table*}

\begin{table*}
    \centering
    \caption{Photometric parameters determined from GP fits: MJD of peak $r-$band light, peak $r-$band magnitude and associated error, the decline rate (\(\Delta\)\textsubscript{m\textsubscript{7}}(r)), and the $g-r$ colour at peak of each object in our sample of confirmed Ca-strong objects. Of the full sample, 10 objects do not have photometric coverage around peak light: SN~2000ds, SN~2001co, SN~2003H, SN~2003dg, SN~2005E, iPTF15eqv, SN~2018gwo, SN~2018lqu, SN~2023fhk, and SN~2024uj. Objects with photometric properties determined using GP fitting have all values and associated uncertainties presented. Objects with time of peak light (and where possible, peak $r-$band magnitude) from their associated publications are labelled with $\dagger$ and the appropriate citation (see footnote). The time of peak light for objects AT~2021zfp, SN~2023fhk, SN~2023ugn, and SN~2024uj were determined through comparison with other Ca-strong objects.}
    \begin{tabular}{lcccc}
    \hline
    Object & MJD\textsubscript{peak}  & M\textsubscript{peak} ($r$) & \(\Delta\)m\textsubscript{7}($r$) & $g-r$ colour \\
     &  & [mag] & [mag] & [mag] \\
    \hline
    SN 2000ds & 51827.0 \footnotesize[$\dagger,1$] & -- & -- & -- \\
    SN 2001co & 52013.0 \footnotesize[$\dagger,1$] & -- & -- & -- \\
    SN 2003H & 52599.0 \footnotesize[$\dagger,1$] & -- & -- & -- \\
    SN 2003dg & 52749.0 \footnotesize[$\dagger,1$] & -- & -- & -- \\
    SN 2003dr & 52779.0 \footnotesize[$\dagger,1$] & -- & -- & -- \\
    SN 2005E & 53377.0 \footnotesize[$\dagger,2$] & -- & -- & -- \\
    SN 2005cz & 53691.0 \footnotesize[$\dagger,3$] & -- & -- & -- \\
    SN 2007ke & 54369.5 \footnotesize[$\dagger,1$] & -- & -- & -- \\
    PTF09dav & 55054.5 $\pm$ 0.3 & $-16.09 \pm 0.04$ & $0.38 \pm 0.07$ & $0.38 \pm 0.08$ \\
    SN 2010et & 55359.4 \footnotesize[$\dagger,4$] & -- & -- & -- \\
    PTF11kmb & 55800.7 $\pm$ 0.2 & $-15.64 \pm 0.01$ & $0.32 \pm 0.01$ & $0.32 \pm 0.02$ \\
    PTF12bho & 55992.7 $\pm$ 0.2 & $-16.15 \pm 0.04$ & $0.54 \pm 0.05$ & $0.54 \pm 0.06$ \\
    PTF15eqv & 57363.0 \footnotesize[$\dagger,5$] & -- & -- & -- \\
    SN 2016hgs & 57692.5 $\pm$ 0.3 & $-15.67 \pm 0.03$ & $0.53 \pm 0.03$ & $0.85 \pm 0.04$ \\
    SN 2016hnk & 57694.5 $\pm$ 0.4 & $-16.99 \pm 0.01$ & $0.37 \pm 0.02$ & $1.28 \pm 0.02$ \\
    SN 2018ckd & 58277.8 $\pm$ 0.3 & $-15.88 \pm 0.02$ & $0.47 \pm 0.03$ & $0.89 \pm 0.04$ \\
    SN 2018dbg & 58307.0 $\pm$ 0.1 & $-16.51 \pm 0.02$ & $0.28 \pm 0.03$ & $0.88 \pm 0.04$ \\
    SN 2018gwo & 58392.0 \footnotesize[$\dagger,6$] & -- & -- & -- \\
    SN 2018kjy & 58461.9 $\pm$ 0.3 & $-15.32 \pm 0.04$ & $0.30 \pm 0.05$ & $0.30 \pm 0.05$ \\
    SN 2018kqr & 58466.9 $\pm$ 0.2 & $-16.66 \pm 0.03$ & $0.14 \pm 0.03$ & $0.78 \pm 0.04$ \\
    SN 2018lqo & 58352.3 $\pm$ 0.1 & $-16.14 \pm 0.02$ & $0.46 \pm 0.03$ & $0.45 \pm 0.06$ \\
    SN 2018lqu & 58370.8 \footnotesize[$\dagger,6$] & -- & -- & -- \\
    SN 2019bkc & 58546.98 $\pm$ 0.1 & $-17.34 \pm 0.03$ & $2.16 \pm 0.12$ & $0.14 \pm 0.02$ \\
    SN 2019ehk & 58615.2 $\pm$ 0.02 & $-15.95 \pm 0.05$ & $0.38 \pm 0.03$ & $1.54 \pm 0.06$ \\
    SN 2019gau & 58630.0 \footnotesize[$\dagger,6$] & -- & -- & -- \\
    SN 2019hty & 58657.9 \footnotesize[$\dagger,6$] & -- & -- & -- \\
    SN 2019ofm & 58723.6 \footnotesize[$\dagger,6$] & -- & -- & -- \\
    SN 2019ouq & 58692.0 \footnotesize[$\dagger,6$] & -- & -- & -- \\
    SN 2019pof & 58748.4 $\pm$ 0.3 & $-16.01 \pm 0.01$ & $0.39 \pm 0.01$ & $0.50 \pm 0.01$ \\
    SN 2019pxu & 58747.8 \footnotesize[$\dagger,6$] & -- & -- & -- \\
    SN 2019txr & 58610.4 $\pm$ 0.4 & $-16.76 \pm 0.04$ & $0.29 \pm 0.05$ & $1.10 \pm 0.04$ \\
    AT 2021zfp & 59479.0 & -- & -- & -- \\
    SN 2021acnz & 59527.0 $\pm$ 0.1 & $-16.27 \pm 0.01$ & $0.30 \pm 0.02$ & $0.50 \pm 0.02$ \\
    SN 2021gno & 59307.2 $\pm$ 0.0 & $-15.14 \pm 0.01$ & $0.46 \pm 0.01$ & $0.50 \pm 0.01$ \\
    SN 2021inl & 59320.7 $\pm$ 0.2 & $-14.69 \pm 0.02$ & $0.82 \pm 0.02$ & $0.72 \pm 0.02$ \\
    SN 2021sjt & 59416.6 $\pm$ 0.1 & $-14.53 \pm 0.01$ & $0.44 \pm 0.02$ & $1.64 \pm 0.02$ \\
    SN 2022nwx & 59765.7 $\pm$ 0.2 & $-15.43 \pm 0.03$ & $0.47 \pm 0.04$ & $0.68 \pm 0.03$ \\
    SN 2022oqm & 59785.1 $\pm$ 0.3 & $-17.33 \pm 0.01$ & $0.30 \pm 0.01$ & $0.46 \pm 0.02$ \\
    SN 2023er & 59937.0 $\pm$ 0.2 & $-17.29 \pm 0.03$ & $0.42 \pm 0.01$ & $0.47 \pm 0.06$ \\
    SN 2023fhk & $\geq$60036.0 & -- & -- & -- \\
    SN 2023ugn & 60232.0 $\pm$ 2.3 & -- & -- & -- \\
    SN 2023xwi & 60274.9 $\pm$ 0.1 \footnotesize[$\dagger,7$] & $-16.8 \pm 0.01$ * & -- & -- \\
    AT 2024aaab & 60617.0 $\pm$ 0.2 & $-16.64 \pm 0.01$ & $0.32 \pm 0.03$ & $1.63 \pm 0.03$ \\
    SN 2024jgw & 60463.1 $\pm$ 0.1 & $-17.26 \pm 0.01$ & $0.53 \pm 0.02$ & $0.89 \pm 0.01$ \\
    SN 2024uai & 60560.8 $\pm$ 0.4 & $-16.52 \pm 0.02$ & $0.51 \pm 0.03$ & $1.02 \pm 0.02$ \\
    SN 2024uj & 60324.0 $\pm$ 1.4 & -- & -- & -- \\
    \hline
    \end{tabular}

    \begin{multicols}{3}
        \footnotesize
        \begin{enumerate}
            \item \citet{kasliwal2012calcium}
            \item \citet{perets2010faint}
            \item \citet{2010Natur.465..326K}
            \item \citet{foley2015kinematics}
            \item \citet{milisavljevic2017iptf15eqv}
            \item \citet{de2020zwicky}
            \item \citet{2025MNRAS.537.1015T}
        \end{enumerate}
    \end{multicols}
    
    \label{tab:photometric_params}
\end{table*}

\begin{table}
    \centering
    \caption{Differences in Akaike ($\Delta$AIC) and Bayesian ($\Delta$BIC) Information Criteria between double- and single-component Gaussian models. Negative values indicate a preference for the double-component model. $|\Delta AIC|\geq5$ and $|\Delta BIC|\geq2$ indicates a strong preference for the double-component model.}
    \label{tab:aic_bic}
    \begin{tabular}{lccc}
    \hline
    Object & Phase [d] & $\Delta$AIC & $\Delta$BIC \\
    \hline
    SN~2000ds   & 14.0  & $-$14.37  & $-$4.83  \\
    SN~2003H    & 99.0  & $-$8.91  & $-$2.01  \\
    SN~2003dg   & 70.0  & $-$16.81 & $-$6.61 \\
    SN~2003dr   & 40.0  & $-$15.74  & $-$6.09  \\
    SN~2005E    & 18.3  & $-$14.59  & $-$8.40  \\
        & 51.3  & $-$13.33 & $-$6.15  \\
    SN~2005cz   & 40.0  & $-$11.68  & $-$1.65  \\
    PTF09dav    & 13.6  & $-$11.75 & $-$4.28 \\
        & 91.6  & $-$8.96 & $-$2.72 \\
    SN~2010et   & 84.6  & $-$13.12  & $-$1.68  \\
    PTF11kmb    & 24.3  & $-$8.95 & $-$2.71  \\
    PTF12bho    & 14.0  & $-$22.72 & $-$11.05 \\
        & 17.0  & $-$20.30 & $-$8.98 \\
        & 52.0  & $-$15.18 & $-$7.08 \\
    PTF15eqv    & 24.0  & $-$17.36  & $-$7.25  \\
        & 42.0  & $-$20.87  & $-$9.91  \\
        & 70.0  & $-$17.92  & $-$7.97  \\
        & 86.0  & $-$17.40  & $-$8.06  \\
        & 129.0 & $-$15.07  & $-$4.35  \\
        & 152.0 & $-$15.88 & $-$4.06 \\
    SN~2016hgs  & 28.2  & $-$8.77 & $-$3.32 \\
      & 58.2  & $-$13.77 & $-$3.41 \\
    SN~2018ckd  & 59.3  & $-$12.69 & $-$2.50 \\
    SN~2018gwo  & 157.0 & $-$16.80  & $-$5.96  \\
    SN~2018kjy  & 26.1  & $-$24.84 & $-$12.82  \\
      & 115.1 & $-$17.34  & $-$5.74 \\
    SN~2018kqr  & 14.0  & $-$4.04 & $-$9.02 \\
    SN~2018lqo  & 51.1  & $-$12.43 & $-$3.62 \\
    SN~2018lqu  & 32.2  & $-$15.04 & $-$4.33 \\
    SN~2019ofm  & 173.4 & $-$5.60  & 0.48  \\
    SN~2019pof  & 149.0 & $-$14.57  & $-$3.19  \\
    AT~2021zfp  & 21.0  & $-$4.83 & $-$2.57  \\
    SN~2022nwx  & 17.0  & $-$12.85  & $-$4.14  \\
      & 30.0  & $-$20.62  & $-$10.37  \\
      & 78.0  & $-$16.39  & $-$5.18  \\
    SN~2022oqm  & 26.0  & $-$9.53 & $-$1.25  \\
      & 60.0  & $-$13.29  & $-$3.45  \\
    SN~2023xwi  & 13.1  & $-$11.90 & $-$4.44 \\
      & 30.1  & $-$10.80 & $-$3.33 \\
      & 86.1  & $-$17.81 & $-$9.32 \\
    SN~2024uj   & 30.0  & $-$16.78  & $-$5.56  \\
       & 31.0  & $-$15.43  & $-$4.52  \\
    SN~2024uai  & 30.0  & $-$18.56  & $-$6.88 \\
    \hline
\end{tabular}
\end{table}

\begin{table}
    \centering
    \caption{Derived envelope mass and radius values for all `present bump' and `possible bump' objects of the 26 objects with photometric observations pre-peak from our total sample.}
    \label{tab:ca_oi6_vels}
    \begin{tabular}{llccc}
        \hline
        SN & Type & $M_e\,[M_\odot]$ & $R_e\,[R_\odot]$ \\
        \hline
        SN 2018kjy & Present & $0.029^{+0.012}_{-0.013}$ & $24^{+12}_{-6}$ \\
        SN 2018lqo & Present & $0.173^{+0.080}_{-0.096}$ & $186^{+47}_{-33}$ \\
        SN 2018lqu & Present & $0.264^{+0.063}_{-0.093}$ & $278^{+99}_{-89}$ \\
        SN 2019ehk & Present & $0.006^{+0.001}_{-0.001}$ & $146^{+22}_{-22}$ \\
        SN 2019hty & Present & $0.40^{+0.17}_{-0.06}$ & $216^{+36}_{-27}$ \\
        SN 2019ouq & Present & $0.054^{+0.059}_{-0.021}$ & $179^{+39}_{-18}$ \\
        SN 2019pof & Present & $0.0196^{+0.0011}_{-0.0010}$ & $111^{+16}_{-16}$ \\
        SN 2019txr & Present & $0.159^{+0.053}_{-0.072}$ & $339^{+72}_{-48}$ \\
        SN 2021gno & Present & $0.0405^{+0.0025}_{-0.0023}$ & $58.4^{+9.3}_{-8.8}$ \\
        SN 2021inl & Present & $0.0100^{+0.0021}_{-0.0026}$ & $81^{+13}_{-12}$ \\
        SN 2021sjt & Present & $0.49^{+0.18}_{-0.13}$ & $19^{+13}_{-13}$ \\
        SN 2021acnz & Present & $0.232^{+0.080}_{-0.054}$ & $322.7^{+3.0}_{-2.1}$ \\
        SN 2022nwx & Present & $0.0250^{+0.0091}_{-0.0078}$ & $94^{+15}_{-15}$ \\
        SN 2022oqm & Present & $0.0495^{+0.0054}_{-0.0051}$ & $1319^{+356}_{-346}$ \\
        SN 2023er & Present & $0.043^{+0.018}_{-0.010}$ & $57.3^{+3.0}_{-2.8}$ \\
        AT 2021zfp & Possible & $\le 0.62$ & $\le 80$ \\
        AT 2024aaab & Possible & $\le 0.013$ & $\le 24$ \\
        SN 2018ckd & Possible & $\le 0.62$ & $\le 56$ \\
        SN 2018dbg & Possible & $\le 0.36$ & $\le 24$ \\
        SN 2018kqr & Possible & $\le 0.013$ & $\le 220$ \\
        SN 2019gau & Possible & $\le 0.62$ & $\le 130$ \\
        SN 2019ofm & Possible & $\le 0.65$ & $\le 110$ \\
        SN 2019ouq & Possible & $\le 0.61$ & $\le 310$ \\
        SN 2023er & Possible & $\le 0.46$ & $\le 24$ \\
        SN 2024jgw & Possible & $\le 0.53$ & $\le 56$ \\
        SN 2024uai & Possible & $\le 0.62$ & $\le 140$ \\
        \hline
    \end{tabular}
\end{table}

\twocolumn
\begin{table}
%\centering
\caption{Velocities and associated errors of the [Ca~II] and [O~I] $\lambda\lambda$6300,6364 features in each spectrum for each confirmed Ca-strong object in our sample. Each line emission velocity is presented only for emission line SNR >3.}
\label{tab:ca_rich_velocities}
    \begin{tabular}{lccc}
        \hline
        Object & $v_{\mathrm{Ca\,II}}\;[\mathrm{km\ s^{-1}}]$ & $v_{\mathrm{O\,I}}\;[\mathrm{km\ s^{-1}}]$ & Phase\tnote{*} [days] \\ \hline
        SN 2000ds & $-245.40 \pm 86.55$ & $-161.12 \pm 192.44$ & 14.0 \\
        SN 2001co & $-303.30 \pm 157.22$ & - & 93.0 \\
        SN 2001co & $-302.86 \pm 131.73$ & - & 104.0 \\
        SN 2001co & $-514.11 \pm 208.70$ & - & 125.0 \\
        SN 2003H & $218.65 \pm 146.06$ & $527.58 \pm 200.18$ & 75.0 \\
        SN 2003H & $-119.77 \pm 99.26$ & $527.58 \pm 200.18$ & 99.0 \\
        SN 2003dg & $-1107.79 \pm 137.60$ & $-1456.31 \pm 220.53$ & 70.0 \\
        SN 2003dr & $444.54 \pm 93.57$ & $529.21 \pm 195.95$ & 40.0 \\
        SN 2005E & $-2139.47 \pm 104.23$ & $180.38 \pm 197.95$ & -3.7 \\
        ... & ... & ... & ... \\
        \hline
    \end{tabular}
 \\The full table can be found in the online supplementary material.

\end{table}

\begin{table}
%\centering
\caption{Sub-classifications of confirmed Ca-strong objects, defined by the line depth of the Si~II~\(\lambda\)6150 absorption feature -- \(a\)\textsubscript{Si~II~\(\lambda\)6150} -- outlined in \cite{sun2017quantitative} and \cite{de2020zwicky}.}
\label{tab:ca_rich_line_depths}
    \begin{tabular}{lccc}
        \hline
        Object & \(a_{\mathrm{Si\,II}\,\lambda\lambda6150} \pm \sigma_{a_{\mathrm{Si\,II}\,\lambda\lambda6150}}\) & Sub-category \\ \hline
        SN 2005E & \(0.093 \pm 0.023\) & Ca-strong Ib/c \\
        SN 2010et & \(0.297 \pm 0.023\) & Ca-strong Ib/c \\
        iPTF11kmb & \(0.078 \pm 0.010\) & Ca-strong Ib/c \\
        iPTF15eqv & \(0.111 \pm 0.009\) & Ca-strong Ib/c \\
        SN 2016hgs & \(0.215 \pm 0.036\) & Ca-strong Ib/c \\
        SN 2016hnk & \(0.639 \pm 0.014\) & Ca-strong Ia \\
        SN 2018gwo & \(0.253 \pm 0.037\) & Ca-strong Ib/c \\
        SN 2018kjy & \(0.137 \pm 0.018\) & Ca-strong Ib/c \\
        SN 2019ehk & \(0.044 \pm 0.027\) & Ca-strong Ib/c \\
        SN 2019gau & \(0.145 \pm 0.006\) & Ca-strong Ib/c \\
        SN 2019kqr & \(0.219 \pm 0.007\) & Ca-strong Ib/c \\
        SN 2019ouq & \(0.063 \pm 0.026\) & Ca-strong Ib/c \\
        SN 2019pof & \(0.181 \pm 0.002\) & Ca-strong Ib/c \\
        AT 2021acnz & \(0.073 \pm 0.002\) & Ca-strong Ib/c \\
        SN 2021gno & \(0.289 \pm 0.026\) & Ca-strong Ib/c \\
        SN 2021sjt & \(0.208 \pm 0.005\) & Ca-strong Ib/c \\
        AT 2022nwx & \(0.200 \pm 0.001\) & Ca-strong Ib/c \\
        SN 2022oqm & \(0.202 \pm 0.002\) & Ca-strong Ib/c \\
        SN 2023er & \(0.108 \pm 0.029\) & Ca-strong Ib/c \\
        SN 2023xwi & \(0.165 \pm 0.036\) & Ca-strong Ib/c \\
        AT 2024aaab & \(0.214 \pm 0.007\) & Ca-strong Ib/c \\
        SN 2024jgw & \(0.091 \pm 0.011\) & Ca-strong Ib/c \\
        SN 2024uai & \(0.121 \pm 0.033\) & Ca-strong Ib/c \\
        \hline
        iPTF09dav & \(0.458 \pm 0.105\) & Ambiguous \\
        SN 2018lqo & \(0.350 \pm 0.022\) & Ambiguous \\
        SN 2018lqu & \(0.256 \pm 0.072\) & Ambiguous \\
        SN 2019ofm & \(0.361 \pm 0.008\) & Ambiguous \\
        SN 2021inl & \(0.177 \pm 0.157\) & Ambiguous \\
        \hline
    \end{tabular}
 \\The following SNe did not have a spectrum in the correct phase range to obtain a measurement: SN~2000ds, SN~2001co, SN~2003dg, SN~2003dr, SN~2003H, SN~2005cz, SN~2007ke, iPTF12bho, SN~2018ckd, SN~2018dbg, SN~2019hty, SN~2019pxu, SN~2019txr, AT 2021zfp, SN~2021inl, SN~2023fhk, SN~2023ugn, and SN~2024uj.

\end{table}

\end{document}